\documentclass[10pt,preprint2]{aastex}

\usepackage{natbib}
\usepackage{graphicx}

\usepackage{graphicx,caption,subcaption}
\usepackage{amsmath}
\usepackage[utf8]{inputenc}
\usepackage [english]{babel}
\usepackage [autostyle, english = american]{csquotes}
\MakeOuterQuote{"}

\usepackage{epstopdf}

\shorttitle{Gaseous Disks of Three HBe Stars: HD\;76534, HD\;114981, and HD\,216629}
\shortauthors{Patel et al.}

\begin{document}


\title{Photoionization Models for the Inner Gaseous Disks of Herbig Be Stars: Evidence Against Magnetospheric Accretion? \footnotemark \footnote{Based on observations obtained at the Canada-France-Hawaii Telescope (CFHT) which is operated by the National Research Council of Canada, the Institut National des Sciences de l''Univers of the Centre National de la Recherche Scientique of France, and the University of Hawaii.}}
\author{P. Patel,\altaffilmark{1} T. A. A. Sigut\altaffilmark{1} and J. D. Landstreet\altaffilmark{2}\\
\affil{Department of Physics and Astronomy, The University of Western Ontario,\\ London, Ontario Canada N6A 3K7}
\email{ppatel54@uwo.ca}}
\altaffiltext{1}{Centre For Planetary Science \& Exploration, The University of Western Ontario, London, Ontario N6A 3K7}
\altaffiltext{2}{Armagh Observatory, Armagh, Co Armagh, UK BT61 9DG}

\begin{abstract}

We investigate the physical properties of the inner gaseous disks of the three, hot, Herbig B2e stars, HD\,76534, HD\,114981 and HD\,216629, by modelling CFHT-ESPaDOns spectra using non-LTE radiative transfer codes. We assume that the emission lines are produced in a circumstellar disk heated solely by the photospheric radiation from the central star in order to test if the optical and near-IR emission lines can be reproduced without invoking magnetospheric accretion. The inner gaseous disk density was assumed to follow a simple power-law in the equatorial plane, and we searched for models that could reproduce observed lines of H\,{\sc i} (H$\alpha$ and H$\beta$), He\,{\sc i}, Ca\,{\sc ii} and Fe\,{\sc ii}. For the three stars, good matches were found for all emission line profiles individually; however, no density model based on a single power-law was able to reproduce all of the observed emission lines. Among the single power-law models, the one with the gas density varying as $\sim 10^{-10}\,(R_{*}/R)^3\, \rm g\,cm^{-3}$ in the equatorial plane of a 25\,R$_{*}$ ($0.78$~AU) disk did the best overall job of representing the optical emission lines of the three stars. This model implies a mass for the H$\alpha$-emitting portion of the inner gaseous disk of $\sim 10^{-9}\,M_*$. We conclude that the optical emission line spectra of these HBe stars can be qualitatively reproduced by a $\approx\,1\;$AU, geometrically thin, circumstellar disk of negligible mass compared to the central star in Keplerian rotation and radiative equilibrium.

\end{abstract}


\addtocounter{footnote}{-1}

\section{Introduction}
\label{sec:intro}

Herbig AeBe (HAeBe) stars are A or B-type, pre-main sequence stars that show optical emission lines, particularly the hydrogen Balmer lines, in their spectra and an infrared (IR) excess in their spectral energy distributions (SEDs); both can be attributed to the presence of circumstellar dust and gas~\citep{Herbig1960, FM1984, WW1998}. HAeBe systems inherit a large amount of angular momentum from the star formation phase, leading the dust and gas to settle into an equatorial disk. HAeBe disks have been the subject of many studies at variety of wavelengths due to their role in star and protoplanetary disk formation (e.g.~\cite{Kraus2015,Wilner2011,Dent2005,Meeus2001,WW1998}). As the disks around HAeBe stars are likely sites of planet formation (e.g.~\cite{Quanz2015,Brittain2013,Quanz2013,Rameau2013,Lagrange2010}), they are of great interest in understanding fundamental disk and planet formation physics. HAeBe stars are the higher mass counterparts to the T Tauri stars\footnote{T Tauri stars are pre-main sequence stars, with masses less than 2.5\,M$_{\odot}$, which show Balmer emission lines in their spectra and an excess in their infrared SEDs, interpreted as radiation from gas and dust in the form of a circumstellar disk.} and represent an important link between the formation of low-mass and high-mass stars. 


The disks of HAeBe stars can be divided into an inner gaseous disk and an outer dusty disk. Due to the high effective temperatures of A and B stars, dust is expected to evaporate near the star. The dust-free, inner gaseous region can be further divided into an inner region of atomic gas and a more distant disk of molecular gas. Emission lines in the optical and near-IR are generally used to study the atomic gas, while near-IR molecular emission lines are used to study the outer molecular gas. As the disk temperature decreases with distance from the star, beyond a certain distance dust is stable and settles into a disk, along with the cool gas. This cool, outer region can also be studied using millimeter and sub-millimeter interferometry (see the review by~\cite{DM2010} and references therein). Several studies using CO overtone emission and millimeter interferometric observation have shown that the gas in these dusty, cool regions follows Keplerian rotation \citep{Ilee2014,MS2000,MS1997}. 

There are various differences amongst the HAeBe stars, as well as between HAeBe stars and T~Tauri stars. While T~Tauri stars are believed to accrete magnetospherically (see \cite{Bouvier2007} and \cite{GomezDeCastro2013} for reviews), there is little direct observational evidence of accretion amongst the early-type HBe stars. Furthermore, if accretion is occurring, it is not clear if the accretion is controlled by large scale magnetic field \citep{Alecian2008,Alecian2013}. Several studies, such as those of~\cite{Vink2002},~\cite{Mottram2007}, and~\cite{Vink2015}, have noted differences in the polarization of Herbig stars: cooler HAe stars show intrinsic H$\alpha$ line polarization, while the hotter HBe stars show line depolarization attributable to dilution by disk H$\alpha$ emission. The authors suggest that this difference reflects different underlying types of accretion, in that linear polarization indicates magnetospheric accretion, while depolarization\footnote{Depolarization occurs when the continuum is more polarized than the photons in the line.} may be explained by disk accretion. 

Recent studies have noted additional differences between HAe and HBe stars in the lines of H$\alpha$, H$\beta$, Br$\gamma$, He\,{\sc i}~$\lambda\,5876$\footnote{All wavelengths are in \AA\ unless otherwise noted.}, He\,{\sc i}~$\lambda\,10830$, Fe\,{\sc ii}~$\lambda\,4924$, and [O\,{\sc i}]$\lambda\,6300$, again suggesting different accretion modes \citep{Mendigutia2011a,Mendigutia2011b,CJK2014,CJK2015}. \cite{Fairlamb2015} studied the UV Balmer excess of the HAeBe stars and showed that magnetospheric models seem unable to reproduce the excess for early-type HBe stars. All of these lines of evidence suggest that an alternative to magnetospheric accretion may be required for the HBe stars and that studying the region close the stellar surface, i.e. the inner gaseous disk, offers a way to shed some light on the mechanism of how and when material from the disk accretes onto HBe stars.

Another class of early-type emission-line stars are the classical Be stars. Classical Be stars are B-type, main-sequence stars that show, or have once shown, one or more Balmer emission lines in their spectrum; they can also display modest amounts of emission in metal lines and and posses an IR excess. However, the IR excess is not attributable to dust but rather to free-free emission from an ionized gas in a circumstellar decretion disk~\citep{Rivinius2013}. Suggestive similarities exist between HBe stars and the classical Be stars in observed features such as polarization level, photometric variability, IR excesses (which can be small for some HBe stars), and H$\alpha$ emission line equivalent width (EW hereafter) distribution \citep{HP1992a,BB2000,Mottram2007}. In particular, \cite{Hillenbrand1992} noted that group III~HBe stars, those with very small IR excesses, closely resemble (and are sometimes confused with!) the classical Be stars. These similarities are likely due to the underlying similarity in the gaseous, circumstellar disks themselves, in spite of the obvious difference of origin (i.e.\ accretion versus decretion). 

This similarity suggests that using computational methods that have been successfully applied to the disks of the classical Be stars may help to better understand and model the physical conditions in the inner gaseous disks of the HBe stars. In this work, we focus on early-type HBe stars as important differences exist between HAe and HBe stars (as stated above), including difficulties in applying magnetospheric accretion models to the early-type HBe stars. The computational modelling tools specifically made for classical Be stars have been very successful in reproducing the observed data (spectra, SEDs, interferometric visibilities), as well as providing a general understanding of the density and temperature structure of classical Be star disks \citep{SJ2007, Tycner2008, Silaj2010,Silaj2014,Sigut2015}.

To study the inner gaseous disk of HBe stars, optical and near-IR emission lines are key diagnostics. Such lines trace not only the structure of the disk (its temperature, density, and velocity field) but also the processes taking place in this region. The strongest emission lines found in the spectra of HBe stars are the hydrogen Balmer lines H$\alpha$ and H$\beta$. Several other emission lines, such as the Ca\,{\sc ii} IR triplet and lines of Fe\,{\sc ii} and O\,{\sc i}, are of moderate strength and appear in emission in the spectrum. H$\alpha$, Ca\,{\sc ii} and Fe\,{\sc ii} lines have been previously used as indicators of wind and chromospheric activity \citep{CK1979,FM1984,HP1992a,CR1998,BC1995}. In contrast, other studies suggest that emission lines such as H$\alpha$ and the Ca\,{\sc ii} IR triplet are non-photospheric in origin and form in a region at some distance from the stellar surface \citep{HP1992a,HP1992b,BC1995}. Previous studies have utilized hydrogen Br$\gamma$ spectro-interferometry for several HAeBe stars with many different types of models, such as magnetospheric models, outflows, and/or keplerian disks (see~\cite{Kraus2015} and reference within). For H$\alpha$, three HAe stars, namely AB Aur, HD 172981, and HD 141529 ~\citep{RP2010,Mendigutia2016}, and one HBe star, MWC 361 (or HD 200775)~\citep{Benisty2013}, have also been studied using spectro-interferometry.

In this paper, we attempt to fit the spectral line profiles of three HBe stars using a circumstellar disk model in which no significant accretion is currently occurring and the disk radiates because it is heated and photoionized by the radiation from the star. This is the second paper in our series investigating inner gaseous disks around early-type HBe stars. \citet{Patel2015}, Paper~I hereafter, modelled optical emission lines of the HB2e star BD+65\,1637 by comparing them to synthetic line profiles computed with the {\sc bedisk} and {\sc beray} non-LTE circumstellar disk codes \citep{SJ2007, Sigut2011}. Acceptable matches to the strengths and shapes of all emission lines considered were found based on a geometrically thin disk with a simple, single power-law parametrization of its density structure. No additional heating of the disk was required beyond the photoionization energy input of the central star. Gas densities of $~\sim\,10^{-10}\,\rm g\,cm^{-3}$ were required in the disk near the star, with the disk extending to $\sim\,25$ stellar radii. However, distinct regions in the disk density parameter space were required to fit the hydrogen and metal lines, leading to the conclusion that the metal lines (Ca\,{\sc ii} and Fe\,{\sc ii}) require a slower density drop-off in the equatorial plane of the disk compared to the hydrogen Balmer lines.

In this current paper, three additional HB2e stars and their inner gaseous disks are investigated using the same numerical codes. Section~\ref{sec:stars} provides background information on the three stars, discusses their fundamental parameters, and outlines the observations and data reduction procedures. Modeling of the line profiles is discussed in Section~\ref{sec:modeling}. The results are discussed in Sections~\ref{sec:results} through \ref{sec:bd65}, and the paper concludes with key findings in Section~\ref{sec:conclusions}.

\section{Stars}
\label{sec:stars}

Three HBe stars of spectral type B2 are considered in this study: HD\,76534, HD\,114981 and HD\,216629.\ The fundamental stellar parameters are adopted from~\cite{Alecian2013} and are given in Table~\ref{table1}. \cite{Alecian2013} derived the effective temperatures by matching synthetic spectra generated by the \texttt{TLUSTY} and \texttt{SYNSPEC} codes \citep{Hubeny1988,HL1992,HL1995} to the observed spectra. The synthetic spectra were generated assuming $\log(g)=4.0$ for all stars. The synthetic spectra were rotationally broadened and used to determine the $v\sin i$ of the stars. Using the stellar luminosities and effective temperatures, the stars were placed in the HR diagram and compared to evolutionary tracks calculated using the CESAM stellar evolutionary code, version 2K~\citep{Morel1997}, to determine the masses and radii.

\begin{table*}[t]
\caption{Adopted Stellar Parameters.}
\label{table1}
\smallskip
\begin{center}{\small
\resizebox{0.95\textwidth}{!}{
\begin{tabular}{lccc} 
\hline \hline
\noalign{\smallskip}
Parameter & HD\,76534 & HD\,114981 & HD\,216629\\
\hline
 \noalign{\smallskip}
\noalign{\smallskip}
T$_{\rm eff}$ (K) & $18000\pm$2000 & $17000\pm$2000 & $19000\pm1000$ \\
\noalign{\smallskip}
Radius  (R$_{\odot}$) & 7.7$\pm$1.6 & 7.0$\pm2.0$ & (6.7) \\
\noalign{\smallskip}
Mass (M$_{\odot}$) & 9.0$\pm0.6$ & 7.9$^{+2.4}_{-1.3}$ & (8.1) \\
\noalign{\smallskip}
Distance (pc) & 870$\pm$80 & 550$^{+260}_{-130}$ & 720$^{+190}_{-150}$ \\
\noalign{\smallskip}
$v\sin i$ ($\rm km\,s^{-1}$) & 68$\pm$30 & 239$\pm13$ & 179$\pm27$ \\
\noalign{\smallskip}
\hline 
\noalign{\smallskip}
\noalign{\smallskip}
	\begin{minipage}{9cm}
		All parameters are from~\cite{Alecian2013}, with the exception of the mass and radius of HD\,216629 which are taken from the B2 average values of Cox (2000).
	\end{minipage}
\end{tabular}}}
\end{center}
\end{table*}

HD\,76534 A (V*~OU~Vel) is a HBe star of spectral type B2e with a visual magnitude of 8.35 located at a distance of 870~pc. It is a member of the Vela R2 association and illuminates the surrounding reflection nebula VdB 24~\citep{Herst1975}. This star is the brighter member of a visual binary. HD\,76534 was first classified as a HBe star by~\cite{FM1984}, and its small infrared excess led it to be classified as group III by \cite{Hillenbrand1992}; group~III HBe stars have small infrared excesses, similar to the classical Be stars. Large variations in the equivalent width (EW) of H$\alpha$, ranging from -14.3~\AA\  to +2.4~\AA, have been reported \citep{OD1997,CR1998,OD1999}. Large scale variations are also seen in the observed polarization of HD\,76534 over time scales of $\sim$1 month~\citep{JB1995}.

HD\,114981 (V* V958 Cen) is a 7.16 visual magnitude star at a distance of 550~pc~\citep{Alecian2013}. It is a B-type star that shows a small IR excess and has been suggested to be a classical Be star. \cite{MB1998} classified HD\,114981 as a B-type star with IR excess typical of Vega-like stars. \citep{CJK2014}. \cite{Hill1970} and \cite{Vieria2003} classified HD\,114981 as HAeBe candidate with spectral type B5 due to H\,$\alpha$ emission; however~\cite{Alecian2013} re-classified it as type B2. \cite{Alecian2013} estimated that HD\,114981 is young with an age of $0.038^{+0.064}_{-0.038}$ Myr. Using the recent analysis, and the evidence presented in the analysis (especially it's young age), HD\,114981 will be treated as a HB2e star.

HD\,216629 (V* IL Cep) is a B2e star of visual magnitude 9.31 located at a distance of 720~pc \citep{Alecian2013}. Several studies, such as \cite{Peter2012} and \cite{Garmany1973}, have associated HD\,216629 with the Cep OB3 association; however, \cite{Alecian2013} argue against this based on the position of HD\,216629 in the H-R diagram. \cite{Wheelwright2010} studied HD\,216629 with spectro-astrometry and found that it is a wide, double-lined, spectroscopic binary. \cite{Alecian2013} also noted variation in He\,{\sc i} lines from one observation to another and indicated that the binary companion might be causing the variations. Double-peaked H$\alpha$ emission has been observed with the EW ranging from -34.5~\AA\ to +4.5~\AA\ \citep{Wheelwright2010,HK2009,Mottram2007,Acke2005,Vink2002}. The EW of H$\beta$ has been reported to be +3.4~\AA\ \citep{Mottram2007}. \cite{Mottram2007} studied the Balmer emission lines of HD\,216629 and failed to detect any depolarization. Because of its binary nature, \cite{Alecian2013} do not assign a mass or radius; therefore, we have adopted average B2 values given by \citet{Cox2000}.

\subsection{Observations}

The current observational data were obtained with the high-resolution ESPaDOnS spectropolarimeter on the Canadian-France-Hawaii Telescope (CFHT). This instrument covers the wavelength range 3700 to 10\,500~\AA\ with a spectral resolution of 65,000.  

\begin{figure*}[ht]
\centering
\includegraphics[width=0.99\textwidth]{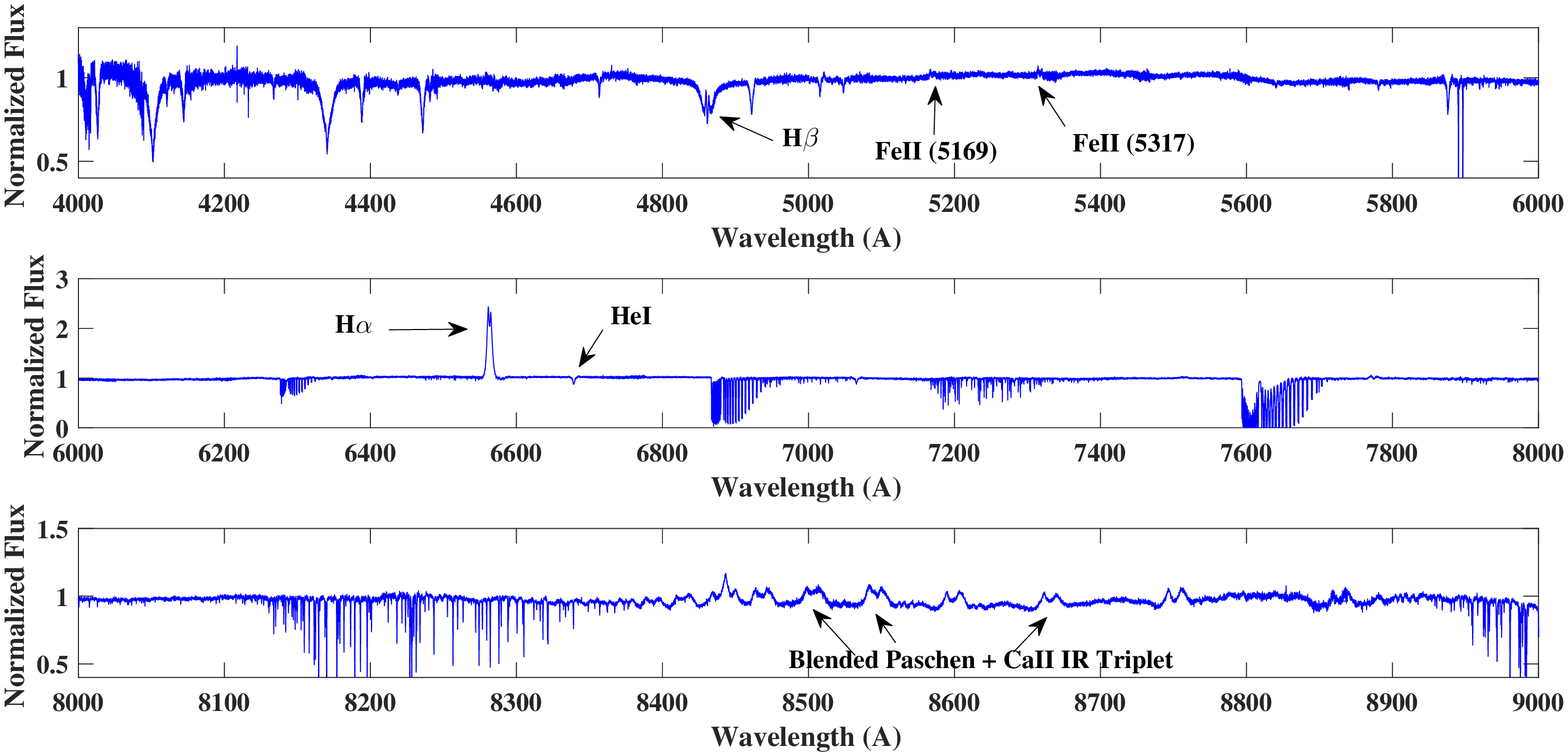}
\caption[The 2005 CFHT ESPaDOnS spectrum of HD\,76534]{The 2005 (HJD\,2453423) CFHT ESPaDOnS spectrum of HD\,76534 from \citet{Alecian2013}. The hydrogen Balmer lines, H$\alpha$ ($\lambda$\,6563) and H$\beta$ ($\lambda$\,4861), the Ca\,{\sc ii} IR-triplet lines ($\lambda$\,8498, $\lambda$\,8542 and $\lambda$\,8662), the Fe\,{\sc ii} lines ($\lambda\,5169$ and $\lambda\,5317$), and He\,{\sc i} ($\lambda\,6678$) are indicated. Note the different vertical scale of each subplot.}
\label{fig:hd76534_spec}
\end{figure*}

The HD\,76534 spectrum, obtained in 2005 (HJD\,2453423), can be seen in Figure~\ref{fig:hd76534_spec}. The peak SNR per pixel is 221 at 7080~\AA. The spectrum contains not only a strong H$\alpha$ emission line but also weaker emission from H$\beta$ and, possibly, from the Ca\,{\sc ii} IR triplet. Weak emission from several Fe\,{\sc ii} multiplets, common in the optical spectra of HBe stars, is also detected. The EWs of all the measured line profiles can be found in Table~\ref{table2}.   

\begin{figure*}
\centering
\includegraphics[width=0.99\textwidth]{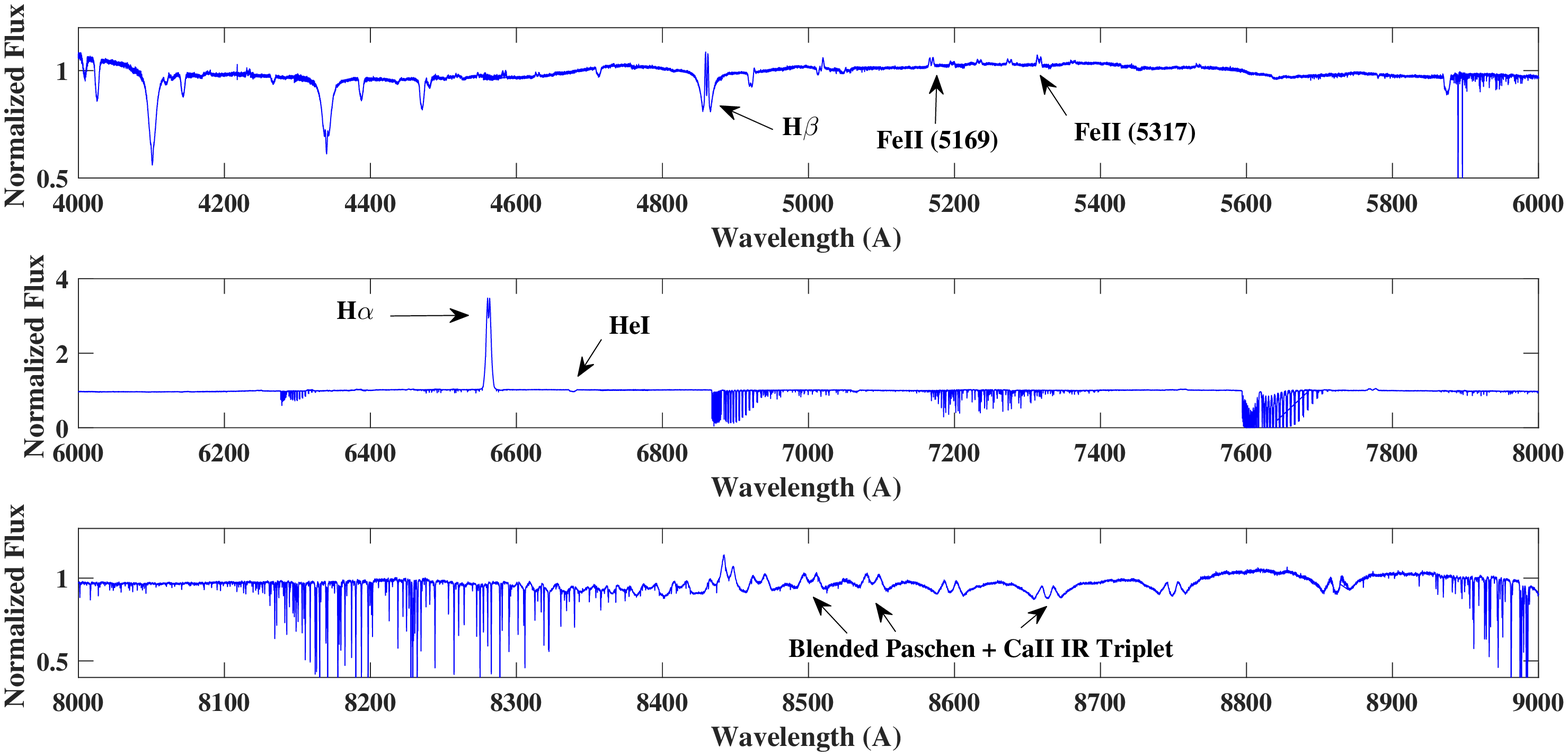}
\caption[The 2006 CFHT ESPaDOnS spectrum of HD\,114981]{The 2006 CFHT ESPaDOnS spectrum of HD\,114981 from \citet{Alecian2013}. Note the different vertical scale of each subplot.}
\label{fig:hd114981_spec}
\end{figure*}

\begin{figure*}[ht]
\centering
\includegraphics[width=0.99\textwidth]{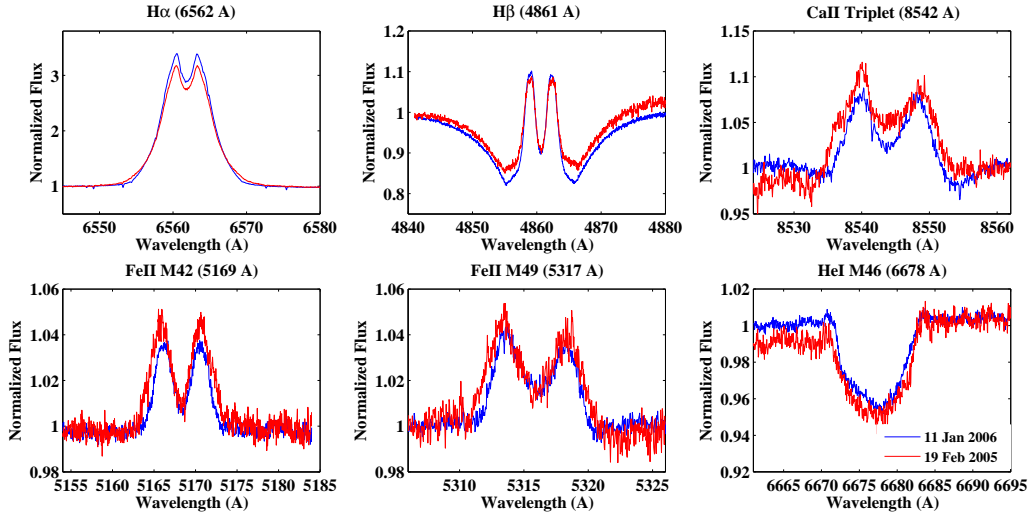}
\caption{Comparison of 2005 (HJD 2453422) and 2006 (HJD 2453748) CFHT ESPaDOnS spectra of HD\,114981.}
\label{fig:hd114981_compare}
\end{figure*}

For HD\,114981, two different spectra are available, one from 2005 (HJD\,2453422) with a peak SNR per pixel of 329 at 5150~\AA, and one from 2006 (HJD\,2453748) with a peak SNR per pixel of 633 at 5150~\AA. It is the 2006 spectrum that is shown in in Figure~\ref{fig:hd114981_spec}. Again, emission lines of H$\alpha$, H$\beta$, possibly Ca\,{\sc ii}, and Fe\,{\sc ii} are detected. A comparison of the two spectra, taken approximately a year apart, can be seen in Figure~\ref{fig:hd114981_compare}, and only small changes are seen. The H$\alpha$ emission is a bit stronger in the 2006 observations, and the H$\beta$ line has shallower wings in the 2005 observations. For the Fe\,{\sc ii} lines, the 2005 profiles are slightly stronger. All emission lines show symmetric, doubly-peaked profiles. He\,{\sc i} $\lambda$6678 is seen in absorption in both spectra and has a very broad profile. As the 2006 spectrum has a higher SNR, we have chosen it for analysis. The EWs of the 2006 line profiles for HD\,114981 can be found in Table~\ref{table2}. 

\begin{table*}[t]
\caption[Measured Equivalent Widths]{Measured Equivalent Widths.}
\label{table2}
\smallskip
\begin{center}{\small
\resizebox{0.95\textwidth}{!}{
\begin{tabular}{lccc} 
\hline \hline
\noalign{\smallskip}
Line and Wavelength (\AA) & \multicolumn{3}{c}{EW (\AA)}\\ \cline{2-4}
    & HD\,76534 & HD\,114981 & HD\,216629\\
\hline
\noalign{\smallskip}
H\,$\alpha$ ($\lambda\,6563$) & -10.9 & -18.6 & -19.7\\
\noalign{\smallskip}
H\,$\beta$ ($\lambda\,4861$) & +4.91 & +4.45 & +2.32\\
\noalign{\smallskip}
Ca\,{\sc ii} ($\lambda\,8542$)$^{a}$ & -1.59 & -0.31 & -1.03\\
\noalign{\smallskip}
Fe\,{\sc ii} ($\lambda\,5169$) & -0.19 & -0.18 & +0.36\\
\noalign{\smallskip}
Fe\,{\sc ii} ($\lambda\,5317$) & -0.60 & +0.20 & -0.17 \\
\noalign{\smallskip}
He\,{\sc i} ($\lambda\,6678$) & +0.51 & +0.40 & +0.66\\
\noalign{\smallskip}
\hline 
\noalign{\smallskip}
\noalign{\smallskip}
	\begin{minipage}{8 cm}
		$^a$ EW after subtraction for Pa15. See Section~2.2.
	\end{minipage}
\end{tabular}}}
\end{center}
\end{table*}

\begin{figure*}
\centering
\includegraphics[width=0.99\textwidth]{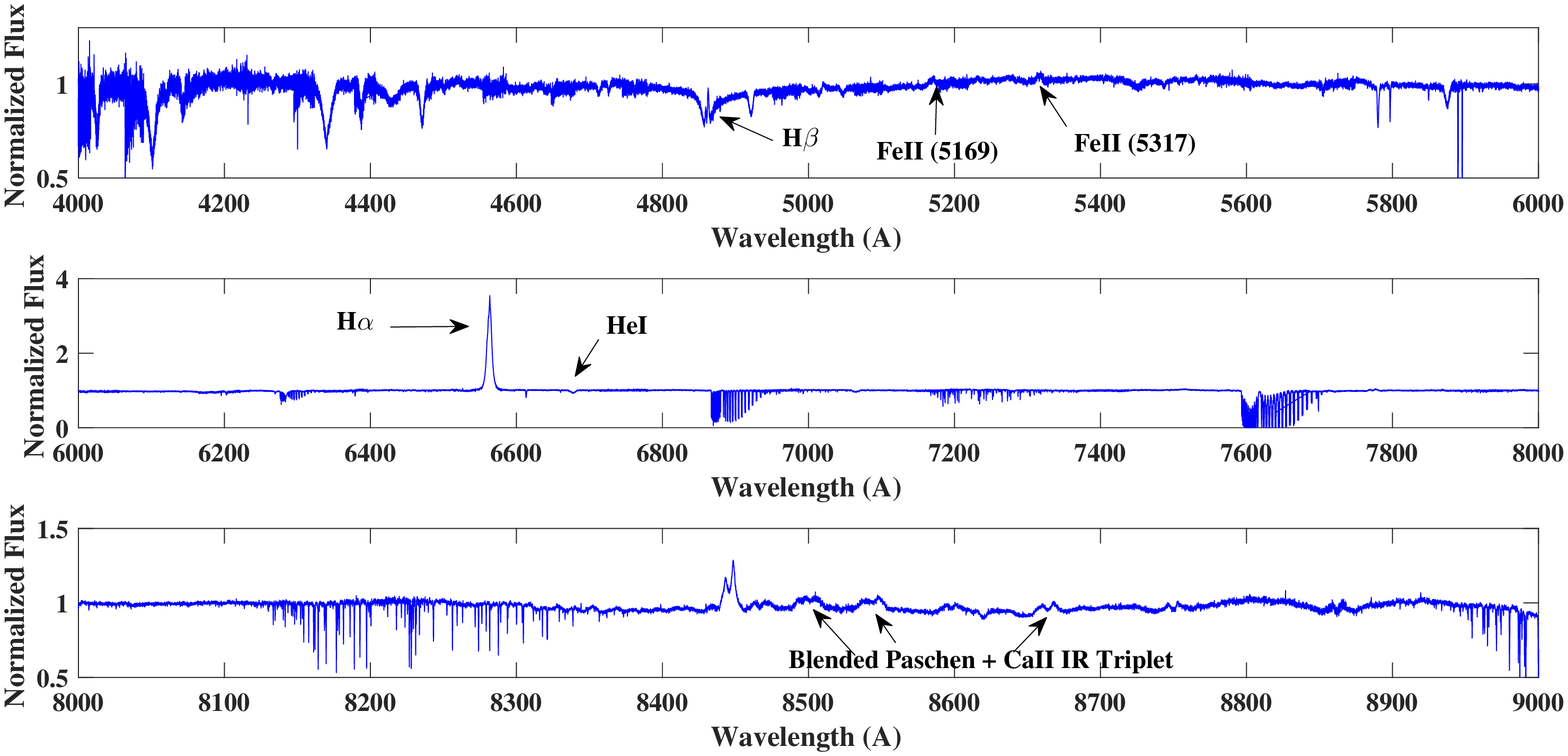}
\caption[The 2006 CFHT ESPaDOnS spectrum of HD\,216629]{The spectrum of HD\,216629 from \citet{Alecian2013}. Note the different vertical scale of each subplot.}
\label{fig:hd216629_spec}
\end{figure*}

The spectrum for HD\,216629 was obtained in 2006 (HJD\,2454077) at peak SNR per pixel of 227 at 7080~\AA\ and can be seen in Figure~\ref{fig:hd216629_spec}. It shows prominent H$\alpha$, H$\beta$, and Ca\,{\sc ii} IR triplet emission lines. Two Fe\,{\sc ii} lines, $\lambda\,5169$ and $\lambda\,5317$, are also present weakly in emission. H$\alpha$ is singly-peaked, while H$\beta$, the Ca{\sc ii} IR triplet, and the Fe{\sc ii} lines are doubly-peaked. H$\beta$ and the Fe{\sc ii} lines show a stronger red (R) component in their doubly-peaked profile compared to the blue (V) component. The EWs of all the line profiles for HD\,216629 can also be found in Table~\ref{table2}. 

For HD\,76534 and HD\,216629, previous observations have suggested variability \citep{OD1997,CR1998,OD1999,Vink2002,Acke2005,Mottram2007,HK2009,Wheelwright2010}; the EW of H$\alpha$ has ranged from -14.3 \AA to +2.4 \AA for HD\,76534 and from -34.5 \AA to +4.5 \AA for HD\,216629. In the current ESPaDOnS spectra, the H$\alpha$ EW for HD\,76534 is -10.9\AA and for HD\,216629, -19.7 \AA; therefore, the current data set for HD\,7634 was obtained when the star was in relatively bright phase, while HD\,216629 was in a more typical phase of its overall variability.

\subsection{Data Reduction}

The spectra for all three stars were continuum normalized using IRAF\footnote{IRAF is distributed by the National Optical Astronomy Observatories, which are operated by the Association of Universities for Research in Astronomy, Inc., under cooperative agreement with the National Science Foundation.} in order to be able to compare the observed line profiles to the synthetic spectra. The individual orders containing the emission lines of interest were separately normalized using various low order (2$^{nd}$ or 3$^{rd}$) polynomials, with either the `Legendre' or `cubic spline' functions available in IRAF.

\begin{figure*}
\centering
\includegraphics[width=0.99\textwidth]{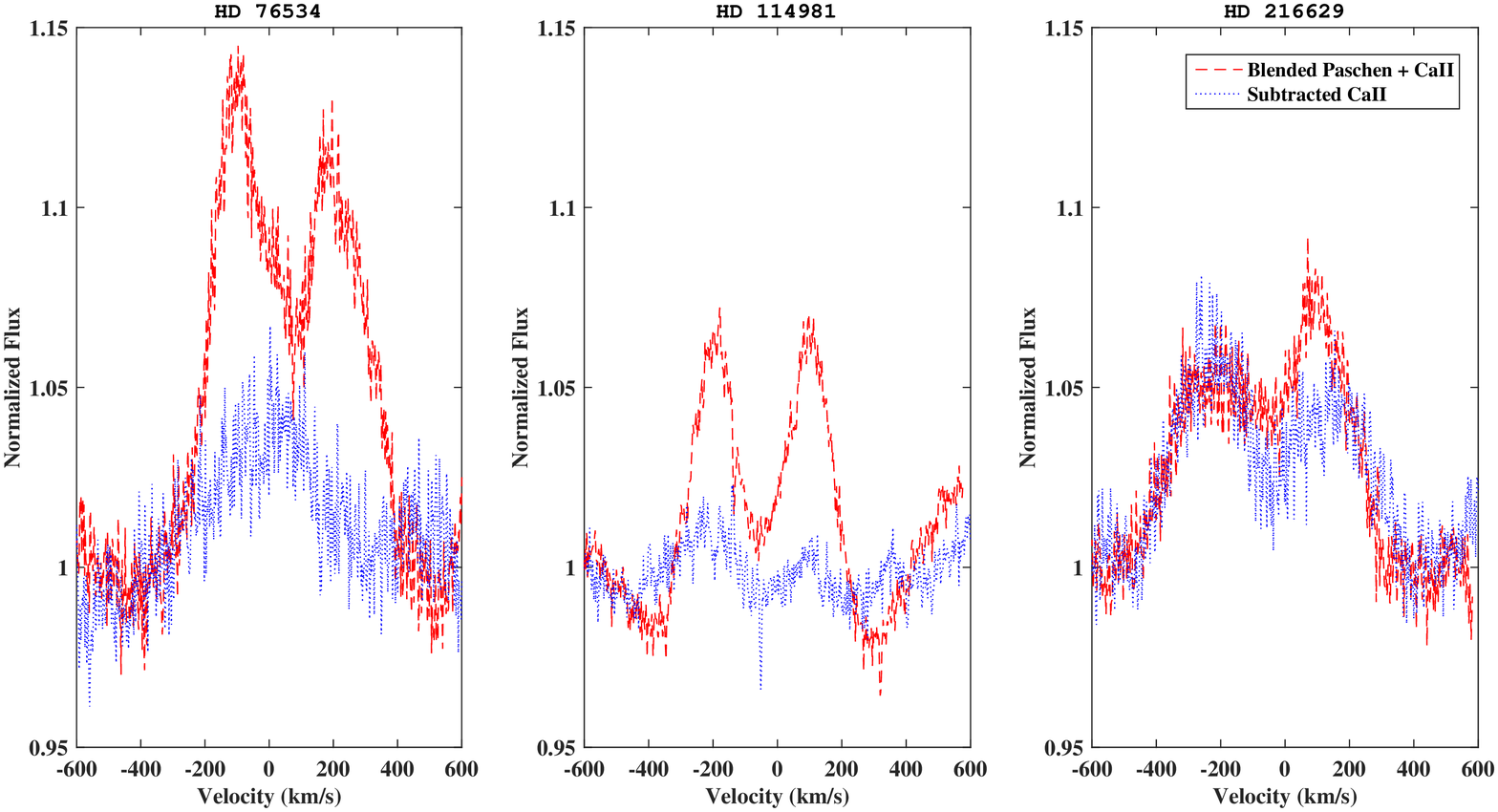}
\caption{The observed Ca\,{\sc ii} $\lambda$\,8542 / Pa15 blend before (red) and after (blue) the subtraction process to remove the Pa15 blend. Each star is shown as a separate panel.}
\label{fig:caIIsubtraction_c4}
\end{figure*}

The three Ca\,{\sc ii} IR triplet lines ($\lambda$\,8498, $\lambda$\,8542, $\lambda$\,8662) are blended with various high-n Paschen lines that needed to be subtracted from the composite profiles in order to extract the unblended Ca\,{\sc ii} line profiles for analysis. In order to remove the contaminating Paschen emission, the two adjacent, unblended Paschen lines were averaged, and this average was subtracted from the composite line profile. For example, to deblend Ca\,{\sc ii} $\lambda$\,8542 from Pa15 $\lambda$\,8543, the average profile of Pa14 $\lambda$\,8596 and Pa17 $\lambda$\,8465 was subtracted from the composite line. This procedure was performed for all the three Ca\,{\sc ii} IR triplets lines for the three stars. An example of the subtraction process, and the result for all three stars, can be seen in Figure~\ref{fig:caIIsubtraction_c4} for the Ca\,{\sc ii} $\lambda$\,8542 emission line. The subtraction process did not yield a significant Ca\,{\sc ii} line profile compared to the continuum noise for two of the three stars, HD\,76534 and HD\,114981. Therefore, Ca\,{\sc ii} was not considered in the analysis of these stars. The Ca\,{\sc ii} line profile for HD\,216629, however, shows a strong, doubly-peaked line profile after subtraction, and this profile was retained in the subsequent analysis. 

\section{Modeling}
\label{sec:modeling}

To compute the synthetic line profiles used to model the observed emission lines, the non-LTE radiative transfer codes {\sc Bedisk} \citep{SJ2007} and {\sc Beray} \citep{Sigut2011}, were used. These codes assume an equatorial, non-accreting, gaseous disk in Keplerian rotation surrounding the central star. The photoionizing stellar radiation field is the sole source of input energy, and the temperature distribution in the disk is found by enforcing radiative equilibrium. The disk is assumed to be axisymmetric about the star’s rotation axis and the disk mid-plane.

Given a user-supplied disk density structure, {\sc bedisk} computes the thermal structure of the disk by enforcing of radiative equilibrium in the gas assumed of solar composition at a large number of grid points in the disk, the net rates of heating and cooling due to nine abundant elements (H, He, C, N, O, Mg, Si, Ca and Fe) over many ionization stages are balanced to determine the radiative equilibrium temperatures. In addition to the temperature structure of the disk, {\sc bedisk} calculates all of the atomic level populations required to compute the gas emissivity and opacity required for radiative transfer. The atomic level populations are found by solving the equations of statistical equilibrium (for more details, see \citet{SJ2007}). 

Line profiles, spectral energy distributions (SEDs), and monochromatic images on the sky can be calculated using the {\sc beray} code \citep{Sigut2011} which solves the equation of radiative transfer along a series of rays ($\approx10^{5}$) that pass through the star+disk system directed at the observer. The disk thermal structure and level populations computed by {\sc bedisk} are used for the radiative transfer calculations. No incident radiation is assumed for the rays that pass through the disk. For rays that terminate on the stellar surface, an appropriate Doppler-shifted, photospheric, LTE line profile was used for the boundary condition. Thus, the synthetic line profiles produced by {\sc Beray} include contributions from both the stellar photosphere and the disk in a consistent way, and the resulting synthetic line profiles, normalized to the predicted continuum, can directly be compared with the observations.

The assumed density structure for the disk was set by two free parameters, a base density $\rho_{0}$ and a power law index $n$, in the equation
\begin{equation}
\rho(R,Z) = \rho_{0} \left(\frac{R_*}{R}\right)^{n} \,
e^{-\left(\frac{Z}{H}\right)^2} \;.
\label{eq:rho}
\end{equation}
Here $R$ and $Z$ are the cylindrical co-ordinates for the disk, $R_{*}$ is the stellar radius, and $H$ is the disk scale height. The disk is assumed to start at the stellar photosphere ($R=R_*$) and extend out to a radius of R$_{disk}$.

The disk was assumed to be in vertical, hydrostatic equilibrium, and in this case, the scale height in Equation~(\ref{eq:rho}) has the form
\begin{equation}
H=\beta(T_{\rm HE}) \left(\frac{R}{R_{*}}\right)^{3/2},
\label{eq:scaleheight}
\end{equation}\\  
where
\begin{equation}
\beta(T_{\rm HE})=\sqrt{\frac{2kT_{\rm HE}\,R_{*}^{3}}{\mu M_{H}\,GM_*}} .
\label{eq:alpha}
\end{equation}\\
Here, $M_{*}$ is the stellar mass, $T_{\rm HE}$ is an assumed, hydrostatic equilibrium temperature used \textit{solely} to fix the vertical structure of the disk,\footnote{It is possible to solve for the vertical density structure of the disk self-consistently with the disk temperatures, $T(R,Z)$, but this self-consistent treatment does not lead to large changes in the predicted emission line profiles \citep{Sigut2009}.} and $\mu$ is the mean molecular weight of the gas, taken to be 0.68. The typical value of $T_{\rm HE}$ in classical Be stars is $T_{\rm HE}\approx 0.6\,T_{\rm eff}$ \citep{Sigut2009}. A thin disk ($H/R\ll1$) is natural consequence of the assumption of vertical hydrostatic equilibrium and the typical radiative equilibrium temperatures found in the disk. However by varying the value of $T_{\rm HE}$, the scale height of the disk can be changed and thicker disks can be produced. Since it is unclear how thick or thin the disks of HBe stars are, both \textit{thin disk models}, achieved by setting the $T_{\rm HE}$ to $0.6\,T_{\rm eff}$, and \textit{thick disk models}, achieved by setting $T_{\rm HE}$ to $5\,T_{\rm eff}$, were considered.

In addition to these two basic disk density models (thick or thin), there is an additional parameter for the adopted disk microturbulence, $\zeta_t$.\footnote{Microturbulence represents the dispersion of an assumed Gaussian distribution of turbulent velocities on scales smaller than unit optical depth. The turbulence acts to broaden the atomic absorption profile and is incorporated as an increase in the Doppler widths of radiative transitions.} The value of microturbulence is set either to zero or the local sound speed in the disk, giving sonic turbulence in the latter case. The models with the microturbulence set to the local sound speed will be referred to as \textit{turbulent disk models}. 

When calculating the synthetic line profiles, an additional model parameter is added to the list: the viewing inclination ($i$) of the disk, defined such that $i=0$ represents a pole-on star / face-on disk. Thus, $\rho_{0}$, $n$, R$_{disk}$, $\zeta_t$, and $i$ are all of the input parameters to be varied for the synthetic line profile calculations. The parameters defining the central B2V star, M$_{*}$, R$_{*}$, T$_{\rm eff}$, are assumed fixed.

The {\sc Bedisk} and {\sc Beray} codes were specifically developed for the gaseous decretion disks of classical Be stars. Many studies (see~\cite{Sigut2015, Silaj2014, SP2013, Grzenia2013, Sigut2011, Silaj2010, Mackay2009, Jones2009, Halonen2008, Jones2008, Tycner2008}) have successfully used the {\sc Bedisk} and {\sc Beray} codes to model Be star gaseous disks by matching observables, such as the hydrogen Balmer lines, Fe\,{\sc ii} lines, IR line fluxes, SEDs, and optical and near-IR interferometric visibilities. Given the apparent similarities between HBe and classical Be stars, these computational codes provide a good starting point towards understanding the inner, gaseous disks around HBe stars using the spectral line profile modelling. In the previous work of Paper~I, the spectrum of the HBe star BD+65\,1637 was successfully modelled using the {\sc bedisk} and {\sc beray} codes.

To model the observed line profiles of the three HBe stars considered in this work, a large grid of synthetic line profiles for disks surrounding a central B2 star were calculated using the parameters given in the Table~\ref{tab:star_param}. The adopted B2 parameters fall within the uncertainty of the stellar parameters given in Table~\ref{table1} for all stars. A wide variety of disk parameters, listed in Table~\ref{tab:model_param}, were explored. Typical values of $\rho_0$  and $n$ for classical Be stars fall in the range of $10^{-12}\,\rm g\,cm^{-3}$ to $10^{-10}\,\rm g\,cm^{-3}$ and 2 to 4, respectively~\citep{Rivinius2013}. However, for HBe stars, an expanded range of model parameters was considered in order to include more massive disks by calculating higher $\rho_0$ and/or lower $n$. Thus, the models were calculated with the base density parameter, $\rho_{0}$, ranging from $10^{-13}$ to $10^{-8}\,\rm g\,cm^{-3}$, and the power-law index, $n$, ranging from $0.5$ to $3.0$ . The size of the disk was chosen to be one of three values, R$_{disk}$=$25$, $50$ or $100$\,R$_{*}$, corresponding to $0.78$, $1.55$, and $3.11$~AU, respectively. All the synthetic line profiles were calculated at four viewing inclinations, which represent the centers of the first four bins of five equal probability bins in a random $\sin i$ distribution: 18$^{\circ}$, 45$^{\circ}$, 60$^{\circ}$ and 75$^{\circ}$. The fifth bin, $i=84^\circ$, was not considered as there is no evidence of shell absorption, i.e.\ deep central reversals in the line profiles caused by disk absorption, in any of the spectra.

\begin{table}[t]
\caption{Adopted stellar parameters for the B2 spectral type.}
\label{tab:star_param}
\smallskip
\begin{center}{\small
\resizebox{0.45\textwidth}{!}{
\begin{tabular}{lc} 
\hline \hline
\noalign{\smallskip}
Parameter & Value\\
\hline
 \noalign{\smallskip}
T$_{\rm eff}$ (K) & 19000\\
log \textit{g} (cgs) & 4.1\\
Radius  (R$_{\odot}$) & 6.7\\
Mass (M$_{\odot}$) & 8.1\\
\noalign{\smallskip}
\hline 
\noalign{\smallskip}
	\begin{minipage}{6 cm}
	Note: values adopted from~\cite{Cox2000}.
	\end{minipage}
\end{tabular}}}
\end{center}
\end{table}

\begin{table}[t]
\caption{Explored model disk density parameters.}
\label{tab:model_param}
\smallskip
\begin{center}{\small}
\resizebox{0.45\textwidth}{!}{
    \begin{tabular}{lc}
    \hline     \hline
\noalign{\smallskip}
    Parameter & Range\\
\noalign{\smallskip}
\hline
\noalign{\smallskip}
Base Disk Density, $\rho_{0}$ $(\rm g\,cm^{-3})$ & $10^{-8}...10^{-13}$\\
Power Law Index, $n$ & $0.5...3.0$\\ 
Inclination, $i$ ($^{\circ}$) & $1
8...75$\\ 
Disk Radius, $R_{disk}$ ($R_{*}$) & $25...100$\\
\noalign{\smallskip}
\hline\
\end{tabular}}
\end{center}
\end{table}

To compare each synthetic profile in the computed library to an observed line profile, a figure-of-merit, ${\cal F}$, was computed. This figure-of-merit was defined as
\begin{equation}
{\cal F} \equiv \frac{1}{N}\,\sum_{i=1}^{N}\,\frac{|F_i^{\rm Mod}-F_{i}^{\rm Obs}|}{F_{i}^{\rm Obs}}, \,
\label{eq:FOM_c4}
\end{equation}
where $F_{i}^{\rm Obs}$ is the observed relative (continuum normalized) flux, $F_i^{\rm Mod}$
is the model relative flux, and the sum is over the $N$ wavelength points spanning the line. A range of small shifts to the observed wavelength scale were also considered, within the errors of the star's radial velocity, in order to best match the observed profile. The smallest value of ${\cal F}$ was deemed to define the best-fit model for that feature, although the ten profiles with smallest ${\cal F}$ values were visually inspected in order to verify a best match. In cases of asymmetric line profiles, i.e.\  V/R (Violet to Red peak) ratios other than 1 (such as H$\beta$ for HD\,216629), the match was made to only one of the peaks, depending on the shape, strength and width of the line profile.

\section{Results}
\label{sec:results}

We will first discuss the best-fit models for each individual line of Table~\ref{table2} for each of the three stars. As will be shown, it is often the case that the best set of model parameters $(\rho_0,n,R_{\rm disk},i)$ will differ between the various lines of the same star. To address this, we consider global fits, that is how well a disk density model based on a single set of these parameters can reproduce the entire emission line spectrum of each star, in Section~\ref{sec:globalfits_c4}, and Section~\ref{sec:uniquenessoffits} discusses the uniqueness of the results. Disk density parameters for all of the individual and global fits for all three stars are summarized in Table~\ref{tab:bestfitmodels_c4}.

\begin{table*}[t]
\caption[Best fit model parameters for individual emission lines and global models for the three HB2e stars]{Individual line and global best-fit model parameters for all three stars. The adopted best-fit, global model for each star is indicated with a check-mark.}
\label{tab:bestfitmodels_c4}
\smallskip
\begin{center}{\small
\resizebox{0.99\textwidth}{!}{
    \begin{tabular}{llccccc}
    \hline
  \hline
  \noalign{\smallskip} \\
    Star & Line & $\rho_{0}$ $(\rm g\,cm^{-3})$ & $n$ & $i$ ($^{\circ}$) & $R_{disk}$ ($R_{*}$) & Model Type\\ 
  \noalign{\smallskip}
  \hline \\
  \noalign{\smallskip}
    HD\,76534 & H$\alpha$ ($\lambda\,6563$) & $7.5\,\times\,10^{-12}$ & 2.5 & 45 & 100 & Thin\\
  \noalign{\smallskip}
     & H$\beta$ ($\lambda\,4861$) & $1.0\,\times\,10^{-11}$ & 3.0 & 45 & 100 & Thick\\
  \noalign{\smallskip}
    & Fe\,{\sc ii} ($\lambda\,5169$) & $1.0\,\times\,10^{-10}$ & 3.0 & 45 & 25 & Thin\\
  \noalign{\smallskip}
   & Fe\,{\sc ii} ($\lambda\,5317$) & $1.0\,\times\,10^{-10}$ & 3.0 & 45 & 25 & Thin\\
  \noalign{\smallskip}
    & Global Fit & $1.0\,\times\,10^{-9}$ & 3.0 & 75 & 25 & Thin \& Turbulent  \\ 
  \noalign{\smallskip}
     & Balmer Line Fit & $1.0\,\times\,10^{-11}$ & 2.5 & 45 & 25 & Thin  \\ 
  \noalign{\smallskip}
     & Metal Line Fit $\surd$ & $1.0\,\times\,10^{-10}$ & 3.0 & 45 & 25 & Thin  \\ 
  \noalign{\smallskip}
 \hline \\
 
 \noalign{\smallskip}
  HD\,114981 & H$\alpha$ ($\lambda\,6563$) & $1.0\,\times\,10^{-12}$ & 2.0 & 60 & 50 & Thick\\
  \noalign{\smallskip}
     & H$\beta$ ($\lambda\,4861$) & $3.2\,\times\,10^{-11}$ & 2.5 & 60 & 100 & Thin\\
  \noalign{\smallskip}
   & Fe\,{\sc ii} ($\lambda\,5169$) & $1.0\,\times\,10^{-10}$ & 3.0 & 45 & 25 & Thin\\
  \noalign{\smallskip}
   & Fe\,{\sc ii} ($\lambda\,5317$) & $1.0\,\times\,10^{-10}$ & 3.0 & 45 & 25 & Thick\\ 
  \noalign{\smallskip}
 & Global Fit $\surd$& $1.0\,\times\,10^{-10}$ & 3.0 & 45 & 25 & Thin  \\ 
  \noalign{\smallskip}
     & Balmer Line Fit & $1.0\,\times\,10^{-11}$ & 2.5 & 45 & 100 & Thin \& Turbulent \\ 
  \noalign{\smallskip}
  & Metal Line Fit $\surd$ & $1.0\,\times\,10^{-10}$ & 3.0 & 45 & 25 & Thin  \\ 
  \noalign{\smallskip}
 \hline \\
 
 \noalign{\smallskip}
  HD\,216629 & H$\alpha$ ($\lambda\,6563$) & $1.0\,\times\,10^{-11}$ & 2.5 & 45 & 25 & Thin \& Turbulent\\ 
\noalign{\smallskip}
     & H$\beta$ ($\lambda\,4861$) & $7.5\,\times\,10^{-11}$ & 3.0 & 45 & 25 & Thin \\
\noalign{\smallskip}
     & Ca\,{\sc ii} IR-triplet ($\lambda\,8542$) & $1.0\,\times\,10^{-10}$  & 3.0 & 60 & 25 & Thick \& Turbulent\\
\noalign{\smallskip}
   & Fe\,{\sc ii} ($\lambda\,5169$) & $1.0\,\times\,10^{-10}$ & 3.0 & 60 & 25 & Thin \\
\noalign{\smallskip}
   & Fe\,{\sc ii} ($\lambda\,5317$) & $1.0\,\times\,10^{-10}$ & 3.0 & 75 & 25 & Thin \& Turbulent \\
   \noalign{\smallskip}
   & He\,{\sc i} ($\lambda\,6678$) & $1.0\,\times\,10^{-12}$ & 1.5 & 75 & 50 & Thick \& Turbulent \\
   \noalign{\smallskip}
      & Global Fit $\surd$ & $1.0\,\times\,10^{-10}$ & 3.0 & 45 & 25 & Thin  \\
  \noalign{\smallskip}
     & Balmer Line Fit & $1.0\,\times\,10^{-11}$ & 2.5 & 45 & 25 & Thin \& Turbulent \\ 
  \noalign{\smallskip}
    & Metal Line Fit $\surd$ & $1.0\,\times\,10^{-10}$ & 3.0 & 45 & 25 & Thin  \\ 
  \noalign{\smallskip}
\hline \\ 

\end{tabular}}}
\end{center}
\begin{minipage}[t]{8cm}
{He\,{\sc i} ($\lambda\,6678$) is a photospheric line and is not used to constrain the disk models except for HD\,216629 (See Section~\ref{sec:linefits_216629} for more details.)}
\end{minipage}
\end{table*}

\subsection{Line Fits: HD\,76534}

\begin{figure*}
\centering
\includegraphics[width=0.99\textwidth]{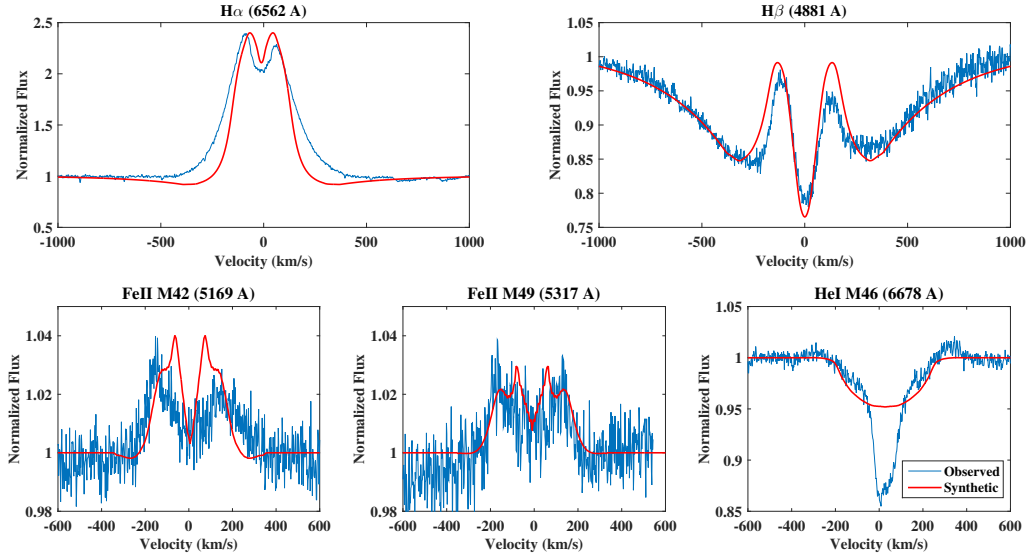}
\caption{The best fit synthetic line profile (red) for each observed emission line of HD\,76534 (blue). The disk density parameters used for each line can be found in Table~\ref{tab:bestfitmodels_c4}, with the exception of He\,{\sc i} as it is fit by a pure photospheric profile.}
\label{fig:hd76534_linefits}
\end{figure*}

The best-fit, synthetic line profiles to the individual observed emission lines for HD\,76534 are shown in Figure~\ref{fig:hd76534_linefits}. All the observed emission lines are reasonably well matched in strength, shape, and width, with the exception of the He\,{\sc i} line, as discussed below. 

The model that best fits the H$\alpha$ line profile is from a $100\,R_{*}$, thin disk model with a base density ($\rho_{0}$) of $7.5\times\,10^{-12}\,\rm g\,cm^{-3}$ and power-law index ($n$) of 2.5 viewed at an inclination of $i=45^{\circ}$. The synthetic line is somewhat narrower at the base when compared to the observed profile, indicating that either slightly more material is required closer to the star and/or the system is viewed at an inclination angle of somewhat more that $45^\circ$. However, the fit was not refined along these lines. 

For H$\beta$, the best-fit model profile is from a $100\,R_{*}$, thick disk model with base density of $1.0\times\,10^{-11}\,\rm g\,cm^{-3}$ and power-law index of 3.0, seen at the same inclination as H$\alpha$, $i=45^{\circ}$. For the two Fe\,{\sc ii} lines, $\lambda$\,5169 and $\lambda$\,5317, the model that fits both observed line profiles is from a 25\,R$_{*}$, thin disk with a base density of $1.0\times\,10^{-10}\,\rm g\,cm^{-3}$ and power-law index of 3.0, again seen at $i=45^{\circ}$. 

Except between the two Fe\,{\sc ii} lines, all of the optimal models for the individual lines differ. As noted before, these best fit models are those with the minimum value of ${\cal F}$. However, there is a range of other disk density models that fit each line profile nearly as well. The range of fitting models for each feature and the implications will be discussed in Section~\ref{sec:uniquenessoffits}.

As for HD\,76534's He\,{\sc i} absorption line, it presents a special problem. The observed line profile has a wide base in the wings and a narrow absorption core. Photospheric synthetic profiles were found to well fit only either the wide wings or the narrow absorption core, depending upon the viewing inclination, with $i=18^{\circ}$ models matching the narrow core and $i=45^{\circ}$ models matching the broad wings. None of the models reproduced the observed line profile as whole. Absorption in the line core by the disk seems an unlikely explanation as absorption is not seen in the cores of the hydrogen lines, and the viewing inclination of the system seems close to $i\approx\,45^\circ$ as based on the other lines. As He\,{\sc i} $\lambda\,6678\,$\AA\ is a transition in helium's singlet spin system, we also do not expect a diffuse, forbidden component to the line profile (as is the case for some of the triplet system). It is important to note that other He\,{\sc i} lines, such as $\lambda\,4713$, $\lambda\,5876$,  and $\lambda\,7065$, show the same broad base and narrow core as seen in He\,{\sc i} $\lambda\,6678$. We have explored  variations in the adopted stellar parameters; models were calculated for $T_{\rm eff}$ increased by 2000\,K (in order to strengthen the helium lines), as well with higher and lower values of $\log\,g$. None of these variations were able to improve the fit to the observed He\,{\sc i} profile. 

\begin{figure}
    \centering
    \includegraphics[width=0.5\textwidth]{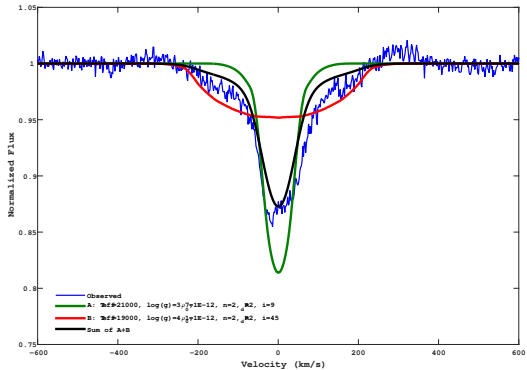}
    \caption{A two component model fit for He\,{\sc i} line profile of HD\,76534. A photospheric line with $v\sin i$ of 222 km/s (red) is fit to the wide base and wings, and a photospheric line with $v\sin i$ of 49 km/s (green) is fit to the narrow absorption core. The sum of the two model lines is also shown (black).}
    \label{fig:heI_twomodels}
\end{figure}

Finally, we investigated the possibility of binarity, postulating that the observed He\,{\sc i} line is actually the sum of two stellar components of very similar spectral types. As mentioned earlier, HD\,76534 is already known to be a visual binary, although the separation is too large for this to be the origin of the addition spectrum. We did manage to produce an acceptable match to the He\,{\sc i} line by combining the two photospheric profiles, i.e.\ a fit to the narrow core with a profile with a $v \sin i$ of 49 km/s and fit the wide base component of the line profile with a model with a $v \sin i$ of 222 km/s. The two line profiles were combined assuming an equal flux contribution from each component, and the composite line profile is illustrated in Figure~\ref{fig:heI_twomodels}. Thus, there is evidence that the HD\,76534 spectrum is actually a composite spectrum of two B stars of nearly the same spectral type. We note that we are unable to detect a systematic velocity offset between the two postulated He\,{\sc i} components; therefore, the binary orbit must be either very wide or the orbital plane is in the plane of the sky. Despite this circumstantial evidence for a composite spectrum, we will retain the assumption that HD\,76534's emission line spectrum is due to a single HBe star in the following analysis.

From Table~\ref{tab:bestfitmodels_c4}, we see that no single model was able to fit all the emission line profiles for HD\,76534. It is likely that a disk density distribution more complex than that of an axisymmetric, single power-law disk, as we have assumed, is necessary. To investigate what range of disk radii contribute to the formation of each line, the cumulative emission for each line was calculated and plotted against radius in the disk for each matching model. To calculate the cumulative disk emission, $C(R)$, a face-on ($i=0^\circ$) synthetic image was produced by {\sc beray} in the light of each emission line using the best-fit disk density model for that line. The intensity was integrated in wavelength over the total width of the line for each image and then this integrated over disk radius as
\begin{equation}
C(R)=2\pi\,\int_{R_*}^{R} I(R^{\prime})\,R^{\prime}\,dR^{\prime} \,.
\label{eqn:c}
\end{equation}
Here $I(R)$ is the wavelength-integrated line intensity at distance $R$, and $R_{*}$ is the stellar radius, assumed to be the inner edge of the disk. Then $C(R)/C(R_{disk})$ can be plotted versus radius of the disk ($R$) to determine the contribution to the total intensity of each line throughout the disk.

\begin{figure*}
\centering
\includegraphics[width=0.9\textwidth]{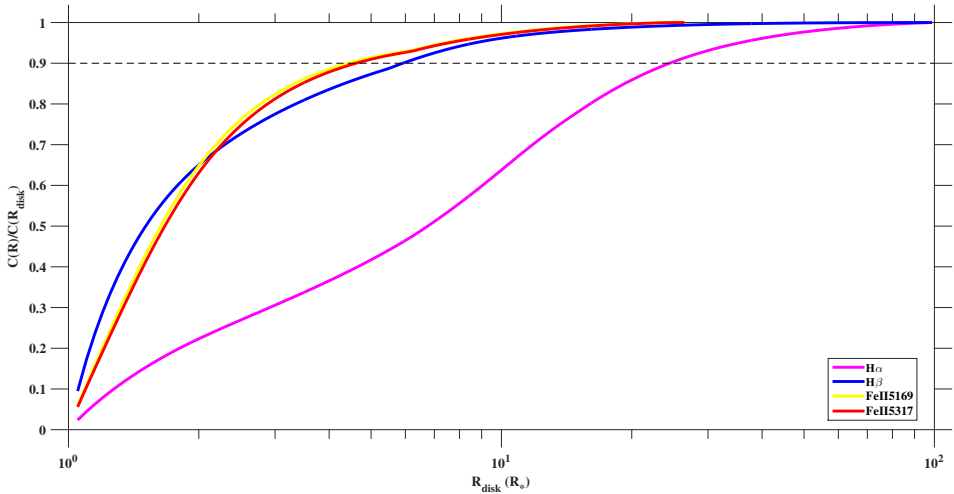}
\caption{The cumulative line emissions (Equation~\ref{eqn:c}) as a function of distance in the disk (in log-scale) from the central star of HD\,76534. The disk density parameters used for each individual line can be found in Table~\ref{tab:bestfitmodels_c4}. Note the the curves for the two Fe\,{\sc ii} lines (red and yellow) overlap. The black dashed line represents the radius inside of which 90\% of the line radiation originates.}
\label{fig:FvsR_hd76534linefits}
\end{figure*}

Figure~\ref{fig:FvsR_hd76534linefits} shows the cumulative emission $C(R)$ for each individual line. It is important to remember that for this figure, the disk density model used to calculate the $C(R)$ for each line is the individual, best fit model for that line given in Table~\ref{tab:bestfitmodels_c4}. As a result, for example,  $C(R)/C(R_{disk})$ reaches one at 100\,R$_{*}$ for H$\alpha$ and H$\beta$, but for Fe\,{\sc ii}, one is reached by 25\,R$_{*}$. While H$\beta$ requires a 100\,R$_{*}$ disk to best reproduce the observed profile, most of the emission comes from the region close to the star. H$\alpha$ also requires extended disk, but 90\% of the emission is produced in the 25\,R$_{*}$ disk near the star. Both the Fe\,{\sc ii} lines are mostly produced within 5\,R$_{*}$ from the stellar photosphere. 

\subsection{Line Fits: HD\,114981}

\begin{figure*}
\centering
\includegraphics[width=0.99\textwidth]{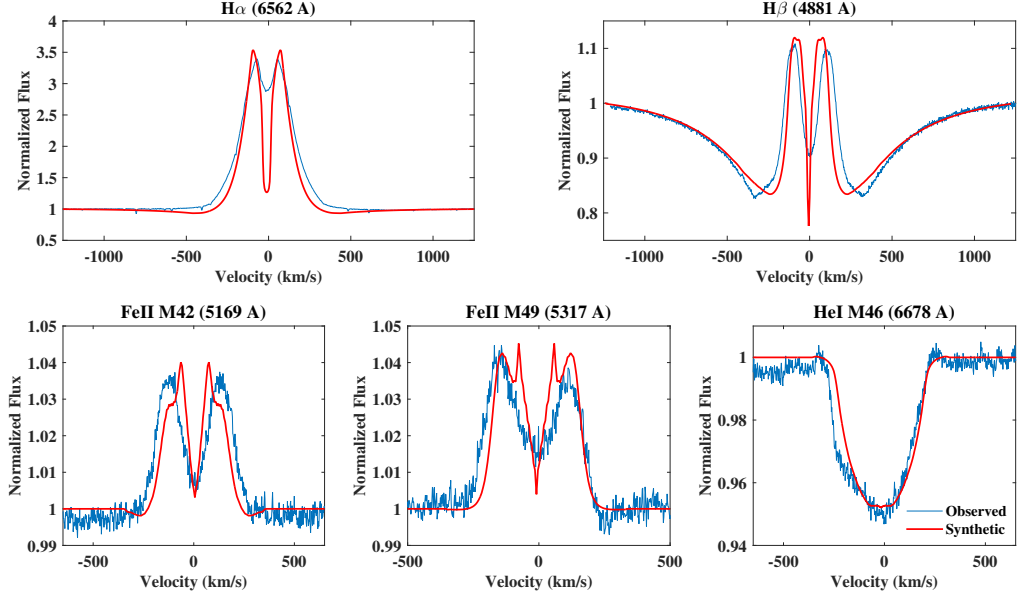}
\caption{The best fitting synthetic emission line (red) for each observed line (blue) of HD\,114981. The disk density parameters used for each line can be found in Table~\ref{tab:bestfitmodels_c4}, with the exception of He\,{\sc i} as it is fit by a photospheric profile. The poor fit to the He\,{\sc i} line is discussed in the text.}
\label{fig:hd114981_linefits}
\end{figure*}

For HD\,114981, the best-fit model parameters, both individual and global, can be found in Table~\ref{tab:bestfitmodels_c4}, and the comparison between the model and observed profiles is shown in Figure~\ref{fig:hd114981_linefits}. For this star, all of the observed lines were individually matched reasonably well in terms of strength and shape by synthetic library profiles. 

The disk density model that best fits the H$\alpha$ line profile is a 50\,R$_{*}$, thick disk with base density ($\rho_{0}$) of $1.0\times\,10^{-12}\, \rm g\,cm^{-3}$ and power-law index ($n$) of 2.0 seen at an inclination of $i=60^{\circ}$. For H$\beta$, the model that best fits the observed line was a 100\,R$_{*}$, thin disk with the base density of $3.2\times\,10^{-11}\, \rm g\,cm^{-3}$ and power-law index of 2.5, again with an $i=60^{\circ}$ inclination. For both Balmer lines, the best-fit synthetic profiles show deeper central absorption compared to the observed lines. A change in the inclination angle, intermediate between 45$^{\circ}$ and 60$^{\circ}$, might improve the fit, but this was not attempted. 

For the two Fe\,{\sc ii} lines, the model that fits both observed profiles well is a 25\,R$_{*}$, base density of $1.0\times\,10^{-10}\, \rm g\,cm^{-3}$ and power-law index of 3.0 when seen at a $i=45^{\circ}$ inclination. The only difference between the fits is that Fe\,{\sc ii} $\lambda\,5169$ requires a thin disk, while Fe\,{\sc ii} $\lambda\,5317$ requires a thick disk. For He\,{\sc i}, the photospheric model that best reproduces the observed line profile has a $v\sin i$ of 222~km/s. The He\,{\sc i} line is matched very well in strength and width.

\begin{figure*}
\centering
\includegraphics[width=0.9\textwidth]{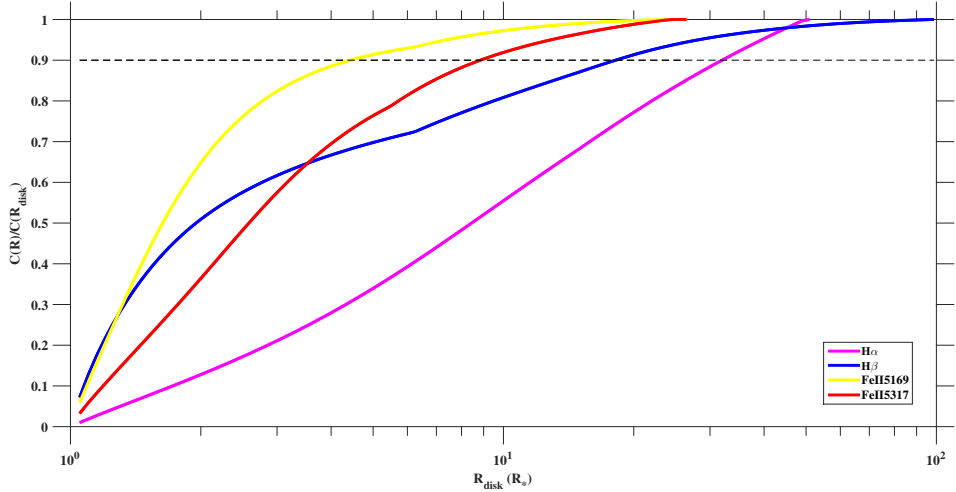}
\caption{The cumulative line emissions (Equation~\ref{eqn:c}) as a function of distance in the disk (in log-scale) from the central star of HD\,114981. The disk density parameters used for each individual line can be found in Table~\ref{tab:bestfitmodels_c4}. The black dashed line represents the radius inside of which 90\% of the line radiation originates.}
\label{fig:FvsR_hd114981linefits}
\end{figure*}

The cumulative disk emission as a function of disk radius for is shown in Figure~\ref{fig:FvsR_hd114981linefits}. Again, it should be remembered that the individual disk density model for each line (Table~\ref{tab:bestfitmodels_c4}) has been used to compute $C(R)/C(R_{disk})$. H$\alpha$ requires region of 50\,R$_{*}$ to reproduce the observed line profile, with most of the emission produced within 30\,R$_{*}$. H$\beta$ is produced throughout the 100\,R$_{*}$ disk; however, 90\% of the emission comes from the inner 20\,R$_{*}$. Finally, Fe\,{\sc ii} $\lambda\,5169$ is formed over a region closer to the star, with most of the emission coming from within 5\,R$_{*}$; Fe\,{\sc ii} $\lambda\,5317$ is formed over a slightly larger range, and 90\% of the emission comes from within 10\,R$_{*}$.  

\subsection{Line Fits: HD\,216629}
\label{sec:linefits_216629}

\begin{figure*}
\centering
\includegraphics[width=0.99\textwidth]{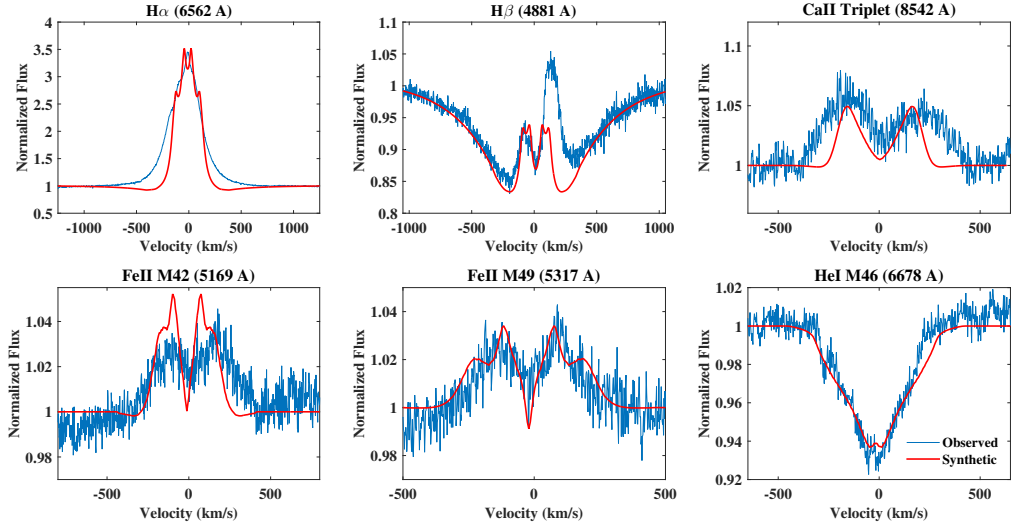}
\caption{The best fit synthetic profiles (red) for each observed emission line line of HD\,216629 (blue). compared to  the observed emission lines (blue). The disk density parameters for each model can be found in Table~\ref{tab:bestfitmodels_c4}.}
\label{fig:hd216629_linefits}
\end{figure*}

The individual, best-fit models for the emission lines of HD\,216629 are illustrated in Figure~\ref{fig:hd216629_linefits}, and the corresponding disk density parameters for each line can be found in Table~\ref{tab:bestfitmodels_c4}. Unlike HD\,76534 and HD\,114981, the Ca\,{\sc ii} infrared-triplet is detected in emission in the spectrum of HD\,216629 after the Paschen line subtraction procedure, and the $\lambda\,8542$ is available for analysis. As can be seen from Figure~\ref{fig:hd216629_linefits}, the hydrogen and iron emission lines are fit reasonably well by the models, but the best Ca\,{\sc ii} IR triplet synthetic line, while fitting the peak emission, is significantly narrower than the observed line profile. 

The synthetic profile that provides the best fit to H$\alpha$ was a 25\,R$_{*}$ thin and turbulent disk with a base density ($\rho_{*}$) of $1.0\,\times\,10^{-11}\,\rm g\,cm^{-3}$ and power-law index ($n$) of 2.5 seen at $i=45^{\circ}$. However, as can be seen in Figure~\ref{fig:hd216629_linefits}, this model gives a small, double-peaked profile in the core, whereas H$\alpha$ is observed to be singly-peaked, and the synthetic line is too narrow at the base near the continuum. The observed H$\beta$ profile shows asymmetric red and blue emission peaks, $R>V$, superimposed on a wide absorption trough. For modelling, the weaker blue emission peak was chosen, and a good fit to both this peak and the absorption trough was found for a 25\,R$_{*}$, thin and turbulent disk with the base density of $1.0\times\,10^{-11}\,\rm g\,cm^{-3}$  and power-law index of 2.5, seen at an inclination of $i=45^{\circ}$. Fitting the red peak instead (not shown in the figure) leads to a larger base density, $1.0\times\,10^{-10}\,\rm g\,cm^{-3}$, and a larger power-law index, $3.0$, but seen at an inclination of $i=60^{\circ}$.

The observed Ca\,{\sc ii} IR triplet ($\lambda\,8542$) emission line is very wide, with a full width of $870\,\rm km/s$ at the base of the profile from the apparent edge of one wing to the other. The best-fit synthetic model for this Ca\,{\sc ii} line is a 25\,R$_{*}$, thick and turbulent disk with a base density of $1.0\,\times\,10^{-10}\,\rm\,g\,cm^{-3}$ and power-law index of 3.0 seen at an angle of $i=60^{\circ}$. The model reproduces the height of the emission peaks, but fails to match the overall width and strength of the line.

Fe\,{\sc ii} $\lambda\,5169$ requires a 25\,R$_{*}$, thin disk with a base density of $1.0\times\,10^{-10}\,\rm g\,cm^{-3}$ and power-law index of 3.0 seen at an inclination of $i=60^{\circ}$. Fe\,{\sc ii} $\lambda\,5317$ line requires the same disk density parameters, but for a thin and turbulent disk seen at a higher inclination of $i=75^\circ$. The synthetic profile for Fe\,{\sc ii} $\lambda\,5169$ is narrower than the observed profile but matches well in strength. The Fe\,{\sc ii} $\lambda\,5317$ synthetic line profile matches well both in shape and strength. 

Unlike HD\,76534 and HD\,114981, the He\,{\sc i} $\lambda\,6678$ line of HD\,216629 required a disk model to fit the observed absorption line profile. The model that best fits the observed line is a 50\,R$_{*}$, thick and turbulent disk with a base density of $1.0\times\,10^{-12}\,\rm g\,cm^{-3}$ and a power law index of 1.5 seen at an inclination of 75$^{\circ}$. The disk is able to produce absorption in the core even at $i=75^\circ$ due to the thick disk and low power-law index for this model. The synthetic line profile shows some central emission which can be attributed to a central, quasi-emission (CQE) feature, consistent with absorption from a circumstellar disk in Keplerian rotation \citep{Han95,Han96}. As noted in Section~\ref{sec:stars}, HD\,216629 shows variation in the He\,{\sc i} line profile from one observation to another. In addition, HD\,216629 is a wide, double-lined spectroscopic binary which is possibly the reason behind the variations.

\begin{figure*}
\centering
\includegraphics[width=0.9\textwidth]{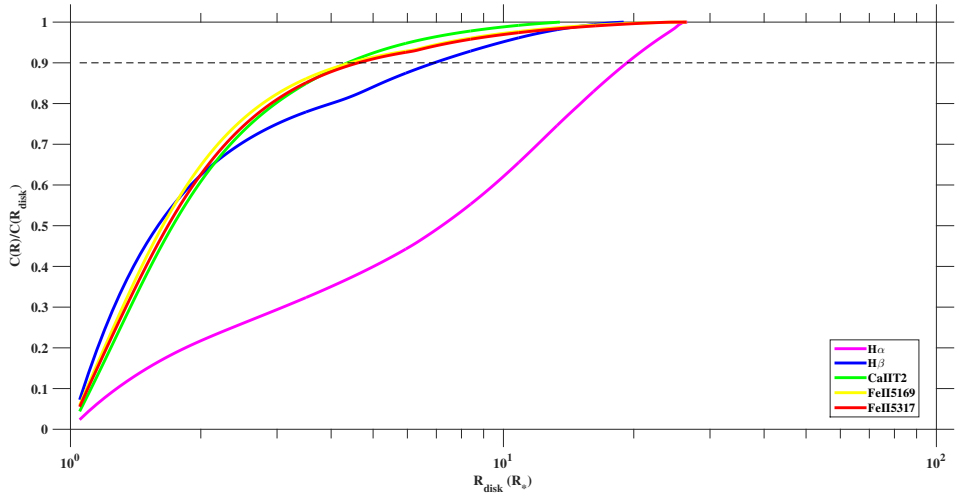}
\caption{The cumulative line emission (Equation~\ref{eqn:c}) as a function of distance in the disk (in log-scale) from the central star of HD\,216629. The parameters used for each individual line can be found in Table~\ref{tab:bestfitmodels_c4}. The black dashed line represents the radius inside of which 90\% of the line radiation originates.}
\label{fig:FvsR_hd216629linefits}
\end{figure*}

A plot of $C(R)/C(R_{disk})$ versus disk radius for HD\,216629 can be found in Figure~\ref{fig:FvsR_hd216629linefits}. Similar to HD\,76534, most of the emission for the Fe\,{\sc ii} lines is produced in the inner most $5\,R_*$ of the disk. For Ca\,{\sc ii}, 90\% of the flux also comes from within $5\,R_*$. For H$\beta$, 90\% of the emission is produced within 7\,R$_{*}$ of the disk, and H$\alpha$ requires a more extended region and is produced within the first 20\,R$_{*}$ of the disk. 

\section{Global Fits}
\label{sec:globalfits_c4}

To identify a unique density structure for each star's inner gaseous disk, the next step was to search for a density model with a single set of parameters $(\rho_0,n,i,R_d)$ and type (thin, thick, turbulent) that could acceptably fit all of the observed line profiles simultaneously. To measure the goodness-of-fit for single disk density model, a weighted sum of all the individual line profile figures-of-merit ${\cal F}$ was computed as
\begin{equation}
{\cal F^{\rm total}} = \sum_{i=1}^{N_l}\, w_{i} {F^{\rm i}} \; ,
\label{eq:FOM_total}
\end{equation}
where $i$ ranges over the lines available, $N_l=6$ if the Ca\,{\sc ii} line is detected in the spectrum with $N_l=5$ otherwise. Initially, the weights $w_i$ were set to one for all lines and such fits will be referred to as the {\it all-line fits}. However, additional weightings were considered: {\it Balmer-line fits}, in which only the hydrogen Balmer lines H$\alpha$ and H$\beta$ were retained in the sum, and {\it metal-line fits}, in which only the Fe\,{\sc ii} and Ca\,{\sc ii} (when available) lines were included in the sum. By examining the results of these various weightings, an adopted density model for the inner, gaseous disk for each star is suggested. The {\it single, best-fit} model was adopted from the above three types of line-fits by visually considering the overall fit of the model line profiles to the observations.

Overall, for the three stars considered, a 25\,R$_{*}$, thin disk with a base disk density ($\rho_{0}$) of $1.0\times\,10^{-10}\,\rm g\,cm^{-3}$ and power-law index ($n$) of $3.0$ seen at $i=45^{\circ}$ is the model that provides the best qualitative fit to all the lines in the spectrum (Table~\ref{tab:bestfitmodels_c4}), given the synthetic model library, and is adopted as the {\it best-fit, single disk} model. It is important to note that this model may not necessarily be the best \textit{all-line fits} model, where an equal weight was given to all the lines. The mass associated with this disk model is $2\times\,10^{25}\,\rm g$ or $10^{-9}\,M_*$ for the three stars.

\begin{figure*}
\centering
\includegraphics[width=0.99\textwidth]{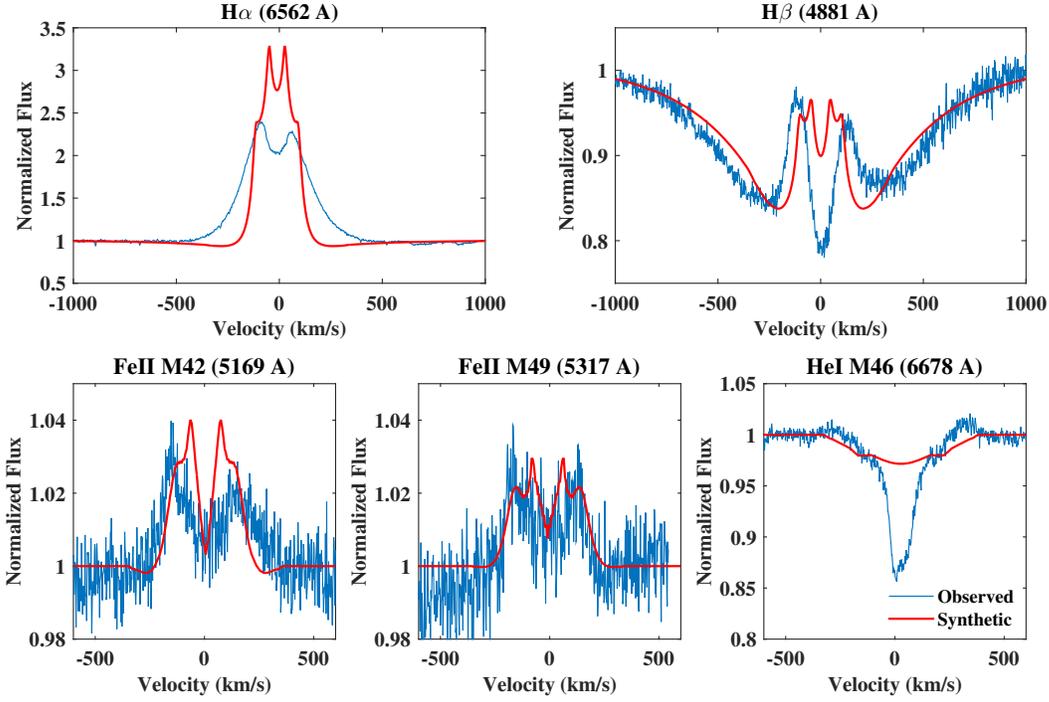}
\caption{Global, metal-line fits for HD\,76534: model profiles (red) are compared to the observed emission lines (blue). The disk model parameters for this fit are $n=3.0$, $\rho_0=1.0\times\,10^{-10}\,\rm g\,cm^{-3}$.}
\label{fig:hd76534_globalfits_metal}
\end{figure*}

Figure~\ref{fig:hd76534_globalfits_metal} shows the synthetic line profiles for the above mentioned model when compared to the observed line profiles of HD\,76534. As can be seen in the figure, the synthetic line profiles for H$\alpha$ and H$\beta$ do not fit the observed line profile very well. They do not possess deep absorption cores because of the $i=45^o$ viewing angle of the model. The Fe\,{\sc ii} lines are much better fit, while the wide base of the He\,{\sc i} line is reproduced. This model was the result of {\it metal-line fits} for HD\,76534.

\begin{figure*}
\centering
\includegraphics[width=0.99\textwidth]{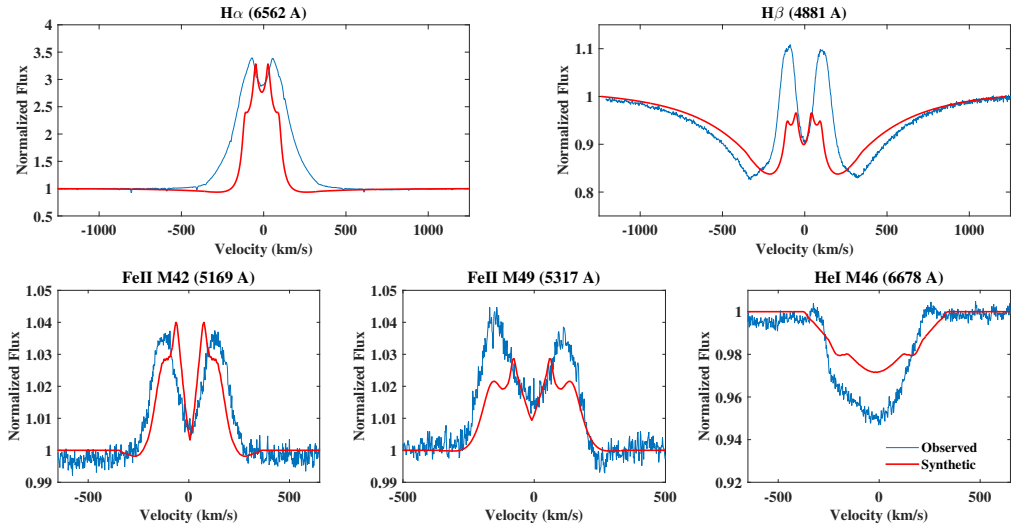}
\caption{Global, all-line and metal-line fit for HD\,114981: model profiles (red) compared to the observed emission lines (blue). The disk model parameters for both weighting, all-line and metal-line, are $\rho_0=1.0\times\,10^{-10}\,\rm g\,cm^{-3}$ and $n=3.0$.}
\label{fig:hd114981_globalfits}
\end{figure*}

Figure~\ref{fig:hd114981_globalfits} shows the {\it best-fit, single disk model} for HD\,114981, reproduced by both the {all-line fit} and {metal-fit} models for this star. As illustrated in the figure, the synthetic profile of H$\alpha$ has the right strength and shape, but is narrower in width than the observed emission line profile. The wings of H$\beta$ are reasonably well reproduced, and the line profile has two peaks and the right shape; however, the synthetic line profile is significantly weaker in the emission peaks than is observed. The synthetic line profile for Fe\,{\sc ii} $\lambda\,5169$ matches to the observed profile well, and for Fe\,{\sc ii} $\lambda\,5317$, the shape of the observed line profile is well reproduced, but it is weaker in strength in the emission peaks. We note that $\lambda\,5317$ shows an asymmetry in the emission peaks that is not seen in $\lambda\,5169$. Lastly, the He\,{\sc i} absorption line profile is weaker than the observed profile.

\begin{figure*}
\centering
\includegraphics[width=0.99\textwidth]{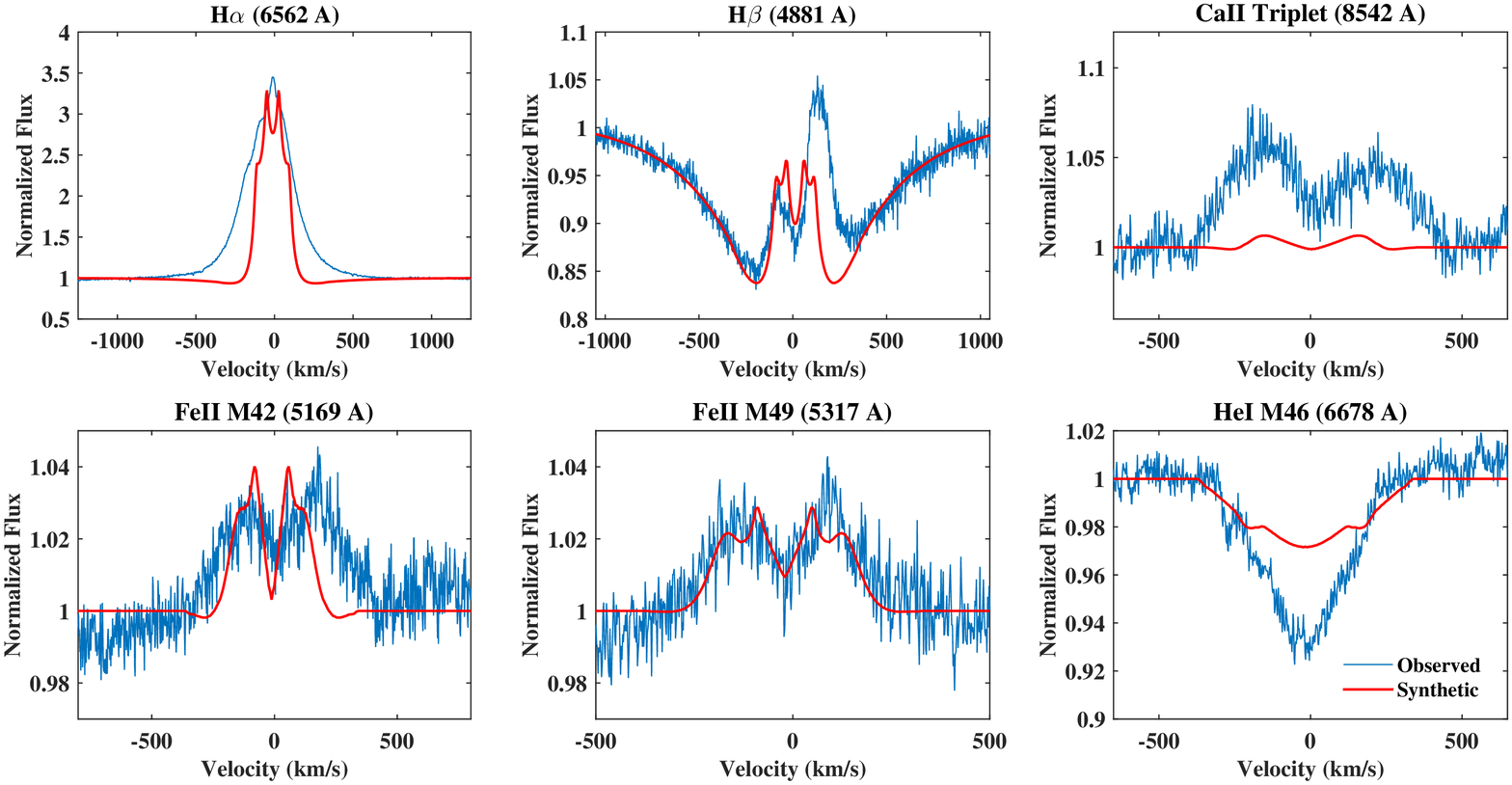}
\caption{Global all-line and metal-line fits for HD\,216629: model profiles (red) compared to the observed emission lines (blue). The disk model parameters for both all-line and metal-line weights are and $\rho_0=1.0\times\,10^{-10}\,\rm g\,cm^{-3}$ and $n=3.0$.}
\label{fig:hd216629_globalfits}
\end{figure*}

Figure~\ref{fig:hd216629_globalfits} illustrates the {\it best-fit, single disk} model for HD\,216629, which was reproduced by both {\it all-line fit} as well as {\it metal-line fit} models. This model reproduces the strength of H$\alpha$ well; however, it is narrower at the base. The H$\beta$ synthetic line profile reproduces the observed blue peak of the line very well; however, the red peak is a poorer match, with the model profile being too narrow and too weak. Unlike HD\,76534 and HD\,114981, the Ca\,{\sc ii} line is detected in the spectrum of HD\,216629. The synthetic Ca\,{\sc ii} line of the all-line model is much too weak compared to the observed line. The synthetic line profile for Fe\,{\sc ii} $\lambda\,5169$ is similar in shape and strength, but narrower compared to the observed line profile. On the other hand, the synthetic line for Fe\,{\sc ii} $\lambda\,5317$ matches the observed line quite well. The synthetic He\,{\sc i} line matches the observed line well in width but produces emission in the core of the line that is not observed.

\begin{figure*}
\centering
\includegraphics[width=0.99\textwidth]{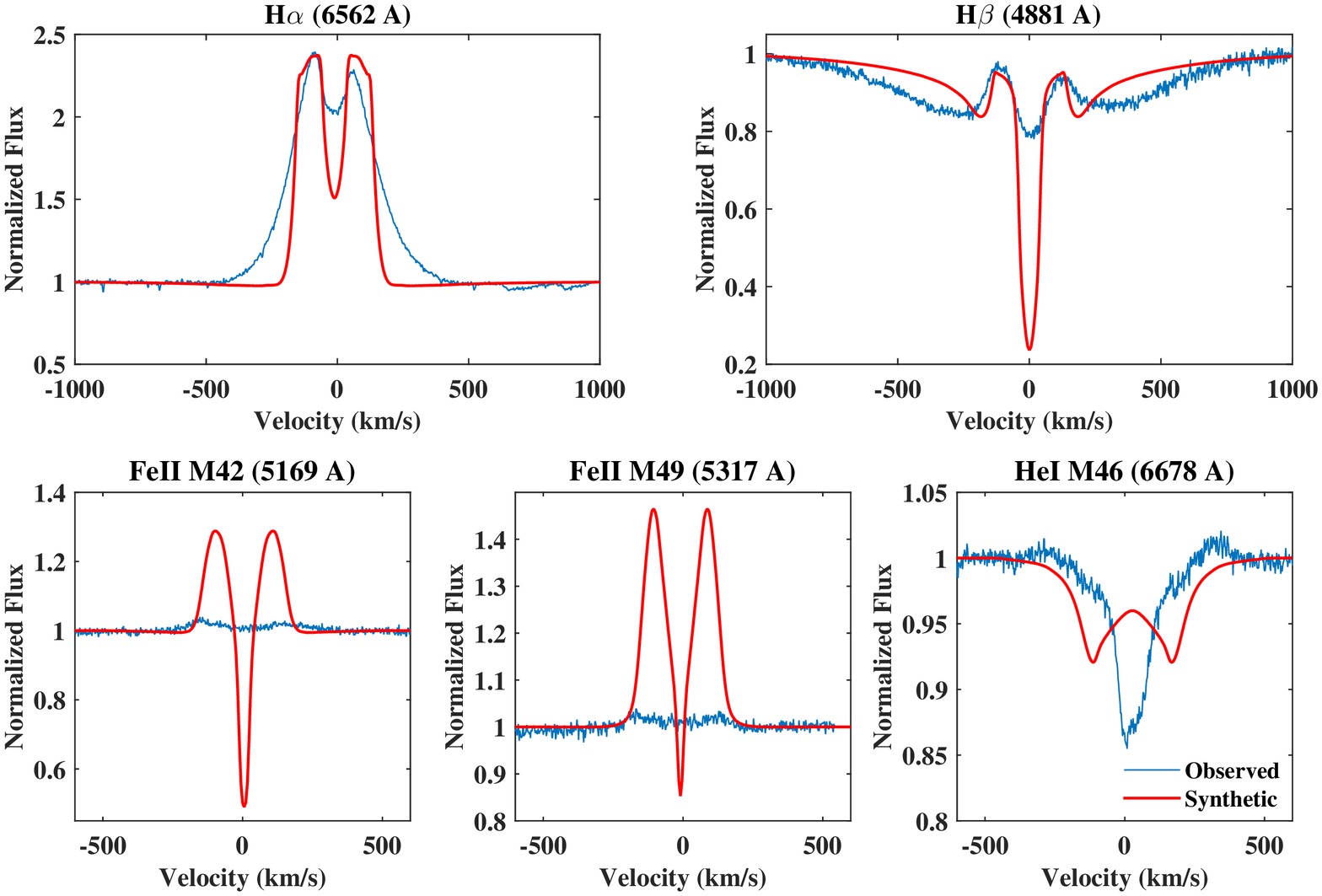}
\caption{Global all-line fits for HD\,76534: model profiles (red) are compared to the observed emission lines (blue). The disk model parameters for this fit are $n=3.0$, $\rho_0=1.0\times\,10^{-9}\,\rm g\,cm^{-3}$; further details can be found in Table~\ref{tab:bestfitmodels_c4}.}
\label{fig:hd76534_globalfits}
\end{figure*}

For HD\,76534, two additional models for the {\it all-line fit} and {\it Balmer-line fit} were found. Figure~\ref{fig:hd76534_globalfits} shows the synthetic line profiles for {\it all-line fits} model as compared to the observed lines. This model has a 25\,R$_{*}$, a thin and turbulent disk with a $\rho_{0}$ of $1.0\times\,10^{-9}\,\rm g\,cm^{-3}$ and $n$ of $3.0$ seen at $i=75^{\circ}$ angle. The H$\alpha$ synthetic line profile has the right strength and shape, but it is narrower at the base compared to the observed profile. As the selected model is seen close to edge-on at $i=75^{\circ}$, the core of the synthetic H$\alpha$ profile shows a deep absorption feature that is not observed. The synthetic line profile for H$\beta$ produces the correct shape for the two peaks; however, it also shows a very deep central absorption core that, like H$\alpha$, is not observed.  Both Fe\,{\sc ii} lines produced by this model are stronger than the observed profiles and also show excessive central absorption. The He\,{\sc i} synthetic line profile is a poor fit; however, the possible two-component nature of this line, illustrated in Figure~\ref{fig:heI_twomodels}, is not accounted for in this model. 

\begin{figure*}
\centering
\includegraphics[width=0.99\textwidth]{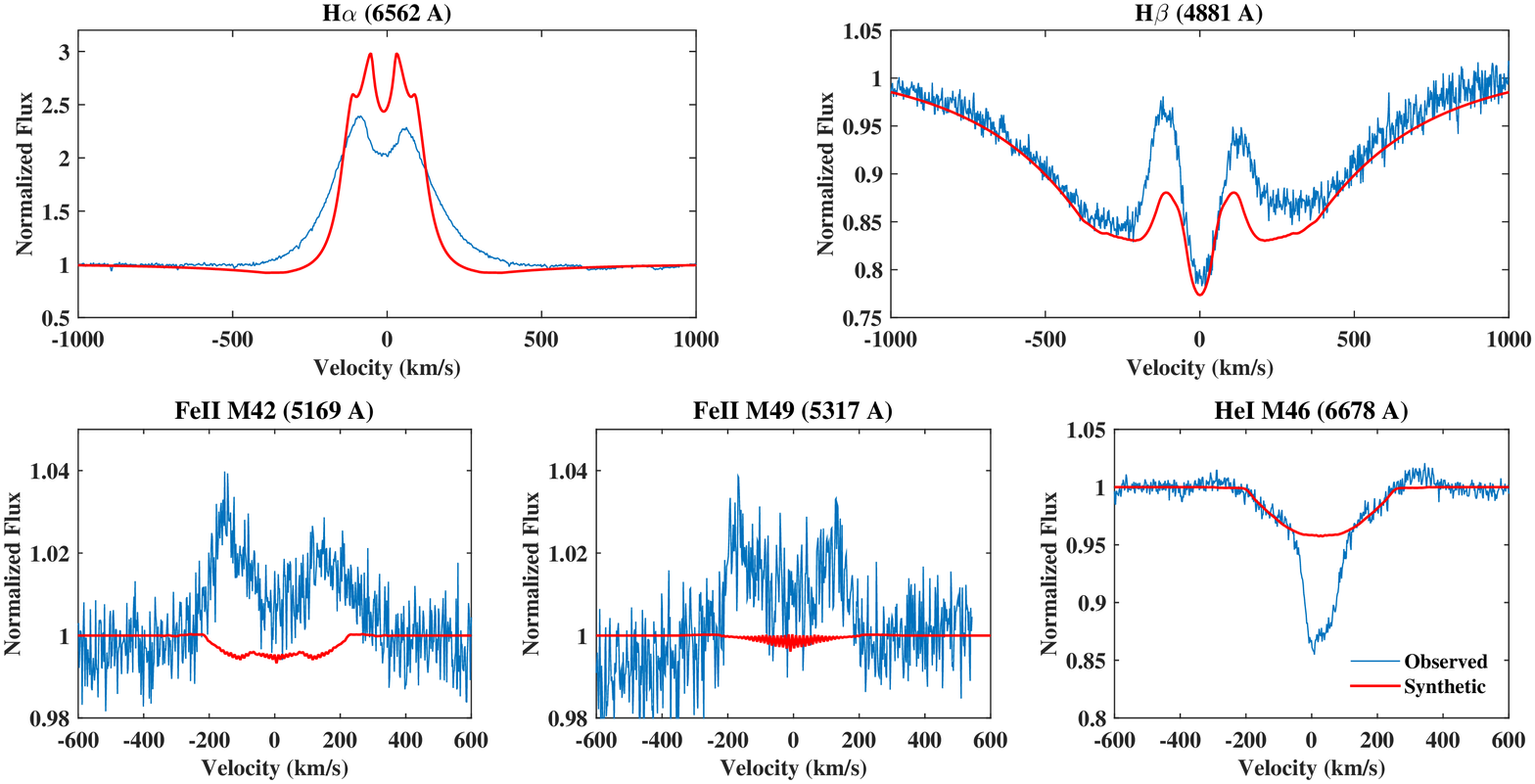}
\caption{Global, Balmer-line fits for HD\,76534: model profiles (red) are compared to the observed emission lines (blue). The disk model parameters for this fit are $n=2.5$, $\rho_0=1.0\times\,10^{-11}\,\rm g\,cm^{-3}$.}
\label{fig:hd76534_globalfits_balmer}
\end{figure*}

Figure~\ref{fig:hd76534_globalfits_balmer} shows the {\it Balmer-line fit} to the observations for HD\,76534. For this model, a 25\,R$_{*}$, a thin disk with a base disk density ($\rho_{0}$) of $1.0\times\,10^{-11}\,\rm g\,cm^{-3}$ and power-law index ($n$) of $2.5$ seen at $i=45^{\circ}$. The model reproduces the observed line profiles for H$\alpha$ and H$\beta$ fairly. The synthetic line profile for H$\alpha$ produces a strong line profile and is also narrower at the base. It is, however, able to reproduce the double-peaked emission profile. For H$\beta$, the synthetic line profile reproduces the core absorption and the shape of the line profile very well, however the strength of the synthetic line is not as strong as the observed. As the metal lines are set to weight-age of zero, the models is not reproducing any of the observed Fe\,{\sc ii} line profile.

\begin{figure*}
\centering
\includegraphics[width=0.99\textwidth]{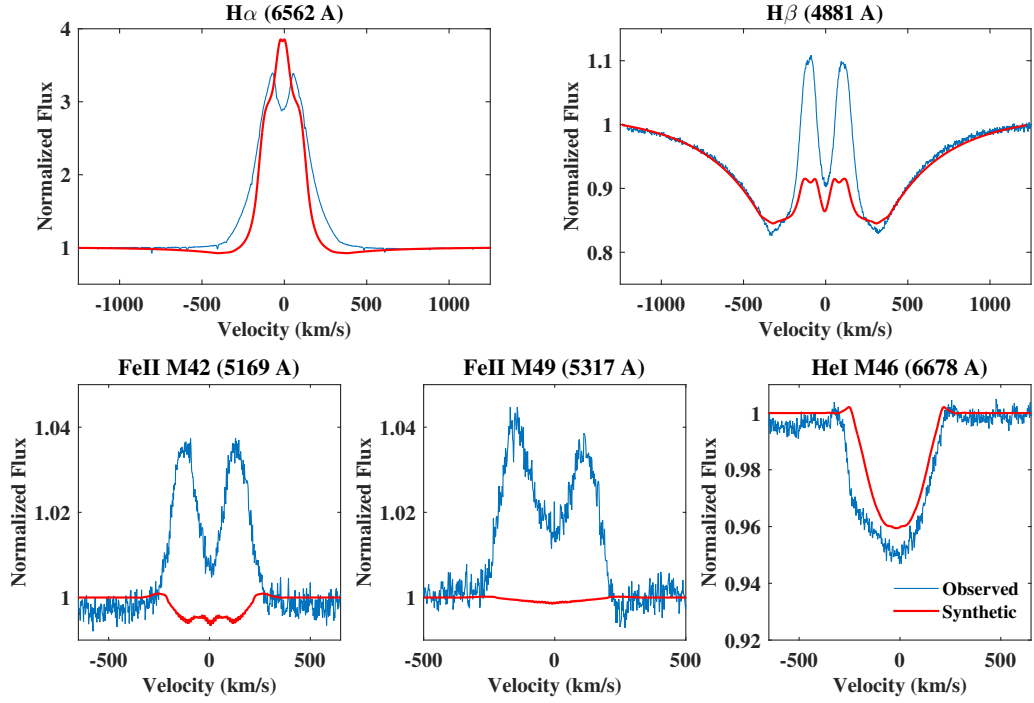}
\caption{Global, Balmer-line fit for HD\,114981: model profiles (red) compared to the observed emission lines (blue). The disk parameters for this fit are $n=2.5$ and $\rho_0=1.0\times\,10^{-11}\,\rm g\,cm^{-3}$.}
\label{fig:hd114981_globalfits_balmer}
\end{figure*}

Figure~\ref{fig:hd114981_globalfits_balmer} illustrates the {\it Balmer-line fit} for HD\,114981. The model parameters can be found in Table~\ref{tab:bestfitmodels_c4} and this model has a $\rho_0$ and $n$ of $1.0\times\,10^{-11}\,\rm g\,cm^{-3}$ and $2.5$, respectively, for 100\,R$_{*}$, a thin and turbulent disk seen at $i=45^{\circ}$. The strength of the H$\alpha$, as well as its overall width, are better reproduced by this model as compared to the all-line fit previously discussed. For H$\beta$, the width of the line and the broad absorption trough is well reproduced, but again the strength of the emission peaks of the synthetic line profile does not match the observed peaks. As for the two Fe\,{\sc ii} lines, the Balmer-line model is an extremely poor match, with this model predicting essentially no Fe\,{\sc ii} emission. For He\,{\sc i} the synthetic line profile is not as the observed, however the shape is reproduced by the model.

\begin{figure*}
\centering
\includegraphics[width=0.99\textwidth]{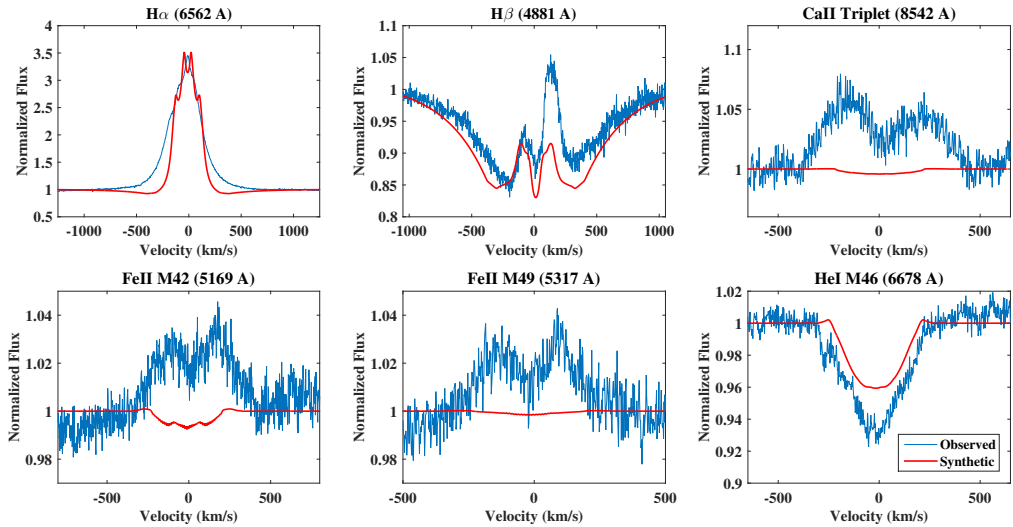}
\caption{Global, Balmer-line fit for HD\,216629: model profiles (red) compared to the observed emission lines (blue). The disk density parameters for this fit are $\rho_0=1.0\times\,10^{-11}\,\rm g\,cm^{-3}$ and $n=2.5$.}
\label{fig:hd216629_globalfits_balmer}
\end{figure*}

Figure~\ref{fig:hd216629_globalfits_balmer} shows the best {\it Balmer-line fit} model for HD\,216629. This model has a $\rho_0$ and $n$, $1.0\times\,10^{-11}\,\rm g\,cm^{-3}$ and $2.5$ respectively, for 25\,R$_{*}$, a thin and turbulent disk seen at $i=45^{\circ}$. The H$\alpha$ synthetic line profile is reasonably reproduced in strength; however, the width of the line at the base is narrower when compared to the observed profile. For the H$\beta$, the strength of the blue peak is reproduced, but the  overall the width of the synthetic line and the size of the red peak is not comparable to the observed profile. Essentially no Fe\,{\sc ii} or Ca\,{\sc ii} emission is predicted by this model. The shape of the observed He\,{\sc i} line profile is reproduced by the model, however it lacks in strength.

\subsection{Summary of the Global Fits}

Remarkably, the adopted, {\it best-fit, single disk} model based on a single set of density parameters for the inner disks of all three stars turns out to be the same: a thin disk with a base density of $\rho_0=1.0\times\,10^{-10}\,\rm\,gm\,cm^{-3}$, a power index of $n=3.0$, and an outer disk radius $R_{d}=25\,R_*$, seen at an inclination of $i=45^{\circ}$. Recall that the only four inclination bins were considered, $i=18$, $45$, $60$, and $75^\circ$, and thus some variation in the viewing inclinations of the three stars is still consistent with the models. The finding of a similar disk model for all three B2 HBe stars is a result of the basic similarity of the strengths and shapes of their emission line spectra, especially the metal lines; this point is further discussed in Section~\ref{sec:bd65}.

\begin{figure*}
\centering
\includegraphics[width=0.9\textwidth]{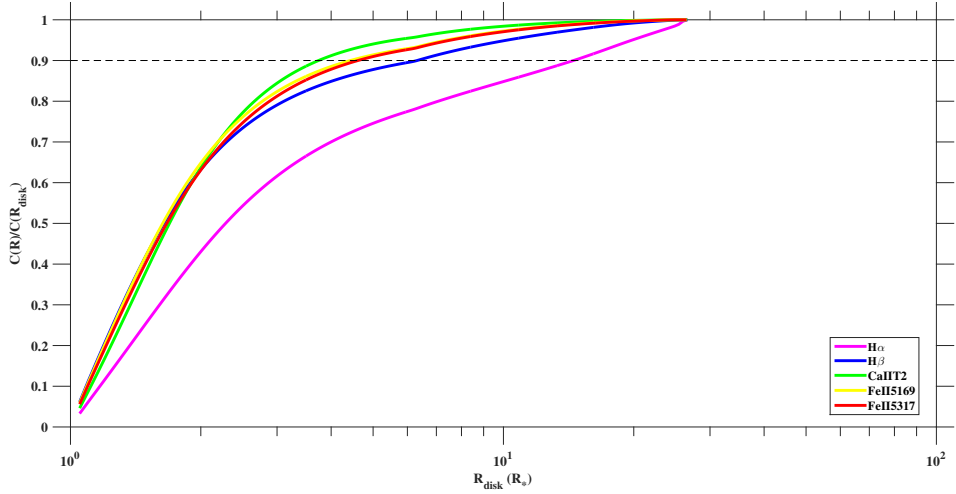}
\caption{The cumulative line emission (Equation~\ref{eqn:c}) as a function of distance in the disk (in log-scale) from the central star for each individual line using the best-fit, global disk density model for HD\,76534, HD\,114981 and HD\,216629. The disk model parameters are $n=3.0$ and $\rho_0=1.0\times\,10^{-10}\,\rm g\,cm^{-3}$; further details can be found in Table~\ref{tab:bestfitmodels_c4}. The black dashed line represents the radius inside of which 90\% of the line radiation originates.}
\label{fig:FvsR_globalfits}
\end{figure*}

To understand where the emission is produced in the disk by this single model, as well as to help in future work to decide how to structure the disk in order to better reproduce the line profiles, the cumulative line emissions, $C(R)/C(R_{disk})$, are plotted as a function of radius for this adopted global model in the Figure~\ref{fig:FvsR_globalfits}. While this plot applies to all three stars considered, HD\,76534, HD\,114981 and HD\,216629, one difference is that Ca\,{\sc ii} emission was detected only in HD\,216629, despite the common model assigned to all three stars. H$\alpha$ is produced in an extended region of the disk, with most of the emission produced within 15\,R$_{*}$. This is followed by H$\beta$, where 90\% of the emission is produced within 7\,R$_{*}$ of the disk. Both the Fe\,{\sc ii} lines follow similar curve and are produced mostly in the inner region of the disk, confined to 5\,R$_{*}$. Given that hardly any emission was seen in the Ca\,{\sc ii} synthetic line profile, the little emission it produced was from the inner 4\,R$_{*}$ of the disk.  



\section{Uniqueness of the Line Fits}
\label{sec:uniquenessoffits}

As discussed in Section~\ref{sec:results}, the minimum value of the figure-of-merit, ${\cal F}_{min}$, defines the best fitting profile. Nevertheless in the complete model library, it is often the case that several models with different combinations of disk parameters $(\rho_0,n,i,R_{\rm disk}$) produce nearly as good fits as ${\cal F}_{min}$ to a given line. To quantify the range of models that provide nearly similar fits to each line, all the models with figure-of-merits ${\cal F}$ satisfying ${\cal F}_{min} \leq {\cal F} \leq 1.25\,{\cal F}_{min}$ have been examined. Table~\ref{tab:25FOM_models_c4} gives the number of such models found for each emission line for each star. The total number of such matching models for each line is seen to vary from one to 17, depending on the line and the star. 

\begin{table*}[th]
\caption{Number of models with ${\cal F}$ satisfying ${\cal F}_{min} \leq {\cal F} \leq 1.25\,{\cal F}_{min}$.}
\label{tab:25FOM_models_c4}
\smallskip
\begin{center}{\small}
   \begin{tabular}{lccc}
   \hline
\hline
   \noalign{\smallskip}
  	Emission Line & HD\,76534 & HD\,114981 & HD\,216629 \\ 
  	\hline
  	\noalign{\smallskip}
 	H$\alpha$ & 1 & 3 & 2 \\ 
	H$\beta$ & 4 & 2 & 2 \\ 
	Ca\,{\sc ii} ($\lambda\,8542$) & -- & -- & 3 \\ 
	Fe\,{\sc ii} ($\lambda\,5169$) & 17 & 1 & 4 \\ 
	Fe\,{\sc ii} ($\lambda\,5317$) & 10 & 1 & 1\\ 
	He\, {\sc i} ($\lambda\,6678$) & -- & -- & 11\\ \noalign{\smallskip}
	\hline
	\end{tabular}
	\end{center}
\end{table*}

In order to illustrate where the ensemble of best-fitting profiles for each line fall in the explored density parameter space, and how well the best fits, ${\cal F}_{min}$, for different lines agree, we have plotted the models satisfying ${\cal F} \leq 1.25\,{\cal F}_{\rm min}$ in the plane defined by the disk density parameters $\rho_0$ and $n$. Note that the other model parameters, $i$ and $R_{disk}$ and type of disk (thin, thick, turbulent, etc...), are not distinguished in such a plot. To graphically illustrate ranges of satisfactory models, we adopt the following conventions: a point is used to illustrate a single model; if there are two selected models, a line connects them, and if there are three models, a triangle is used. If there are more than three models satisfying the condition on ${\cal F}$, an ellipse is used to enclose the points. For He\,{\sc i}, which generally has a photospheric origin and thus does not constrain the disk parameters, an arrow and line of the same color are used to represent the region of density parameters that produce no disk emission. Finally, in these figures, the adopted global disk model for all stars is at the point $(n,\log\rho_0)=(3.0,-10.0)$.

\begin{figure*}
\centering
\includegraphics[width=0.65\textwidth]{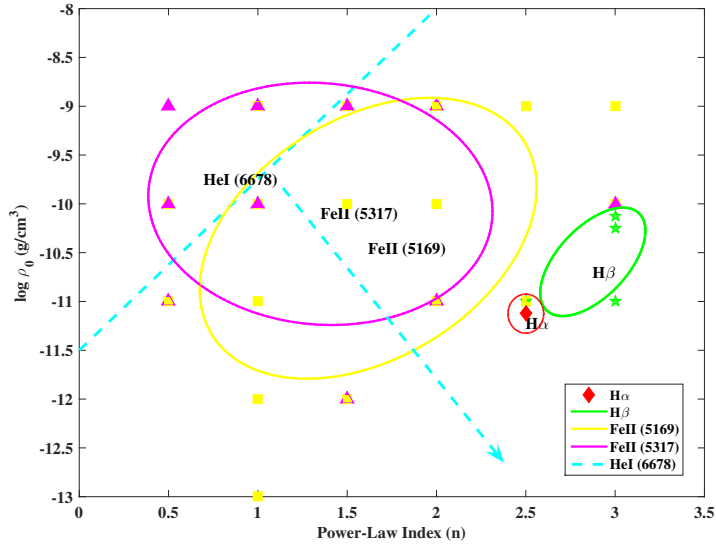}
\caption{Regions of the $(n, \log\rho_{0})$ plane occupied by the modelled lines of HD\,76534 with ${\cal F}\leq 1.25\,{\cal F}_{min}$.}
\label{fig:hd76534_ellipseplot}
\end{figure*}

Figure~\ref{fig:hd76534_ellipseplot} shows the $(n,\log\rho_0)$ plane for HD\,76534. Only some of the models for the different emission lines overlap in the disk density parameter space, illustrating the conclusion of Table~\ref{tab:bestfitmodels_c4} that there is no single set of disk parameters in a single power-law for the disk's equatorial density for which all of the lines can be well-fit. For HD\,76354, two separate regions in parameter space can be seen, one in which the Fe\,{\sc ii} lines are best fit and another where the fits to the two hydrogen Balmer lines are found. The Fe\,{\sc ii} lines select a large region in parameter space corresponding to ${\cal F} \leq {\cal F}_{min}$, whereas the Balmer lines select a much smaller region that is of comparable base density but larger power-law index.

\begin{figure*}
\centering
\includegraphics[width=0.65\textwidth]{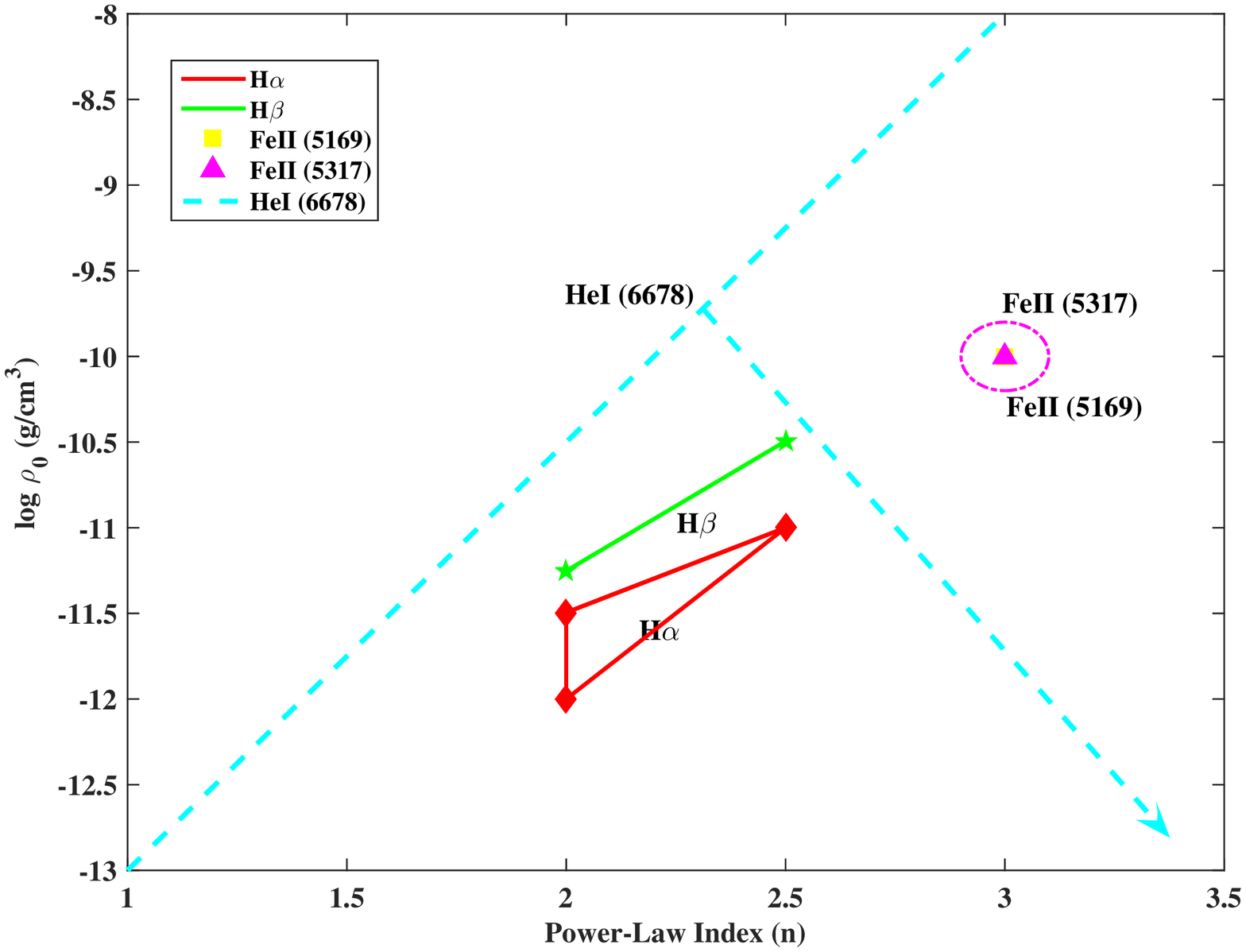}
\caption{Regions of the $(n, \log\,rho_{0})$ plane occupied by the modeled lines of HD\,114981 with ${\cal F}\leq 1.25\,{\cal F}_{min}$.}
\label{fig:hd114981_ellipseplot}
\end{figure*}

\begin{figure*}
\centering
\includegraphics[width=0.65\textwidth]{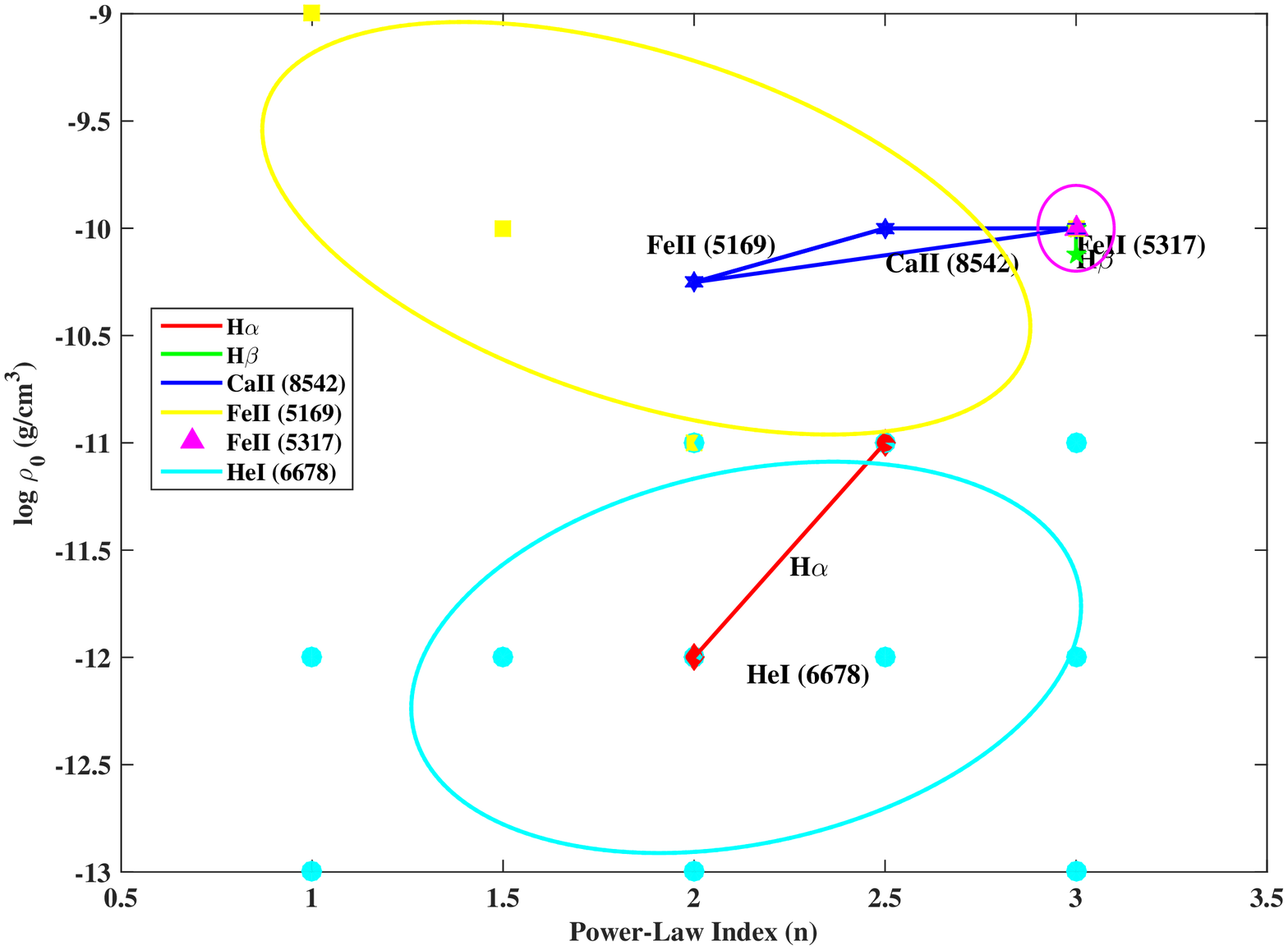}
\caption{Regions of the $(n, \log\rho_{0})$ plane occupied by the modelled lines of HD\,216629 with ${\cal F}\leq 1.25\,{\cal F}_{min}$.}
\label{fig:hd216629_ellipseplot}
\end{figure*}

The plot illustrating the $(n,\log\rho_0)$ plane for HD\,114981 can be seen in Figure~\ref{fig:hd114981_ellipseplot}. The regions selected are much smaller than for HD\,76534. Again the Balmer and Fe\,{\sc ii} lines select quite different regions, although in this case, the Balmer lines favor both lower base density and power-law index as compared to the Fe\,{\sc ii} lines. Finally, Figure~\ref{fig:hd216629_ellipseplot} illustrates the models for HD\,216629. For this star, H$\alpha$ requires somewhat smaller base densities than the region selected by the other lines. HD\,216629 is the single star which includes the Ca\,{\sc ii} IR triplet which favors the region selected by H$\beta$ and the Fe\,{\sc ii} lines. HD\,216629 also the only star in the study that requires disk models to fit the observed line profile for the He\,{\sc i} line profile, rather than a photospheric profile. The He\,{\sc i} covers regions of low density at variety of power law indices, which overlaps with only the H$\alpha$ profile. 

Finally note that the adopted global model for all three stars would fall at the point $\log\rho_0=-10$ and $n=3$ in these plots. In general, this is the region selected by the metal lines Fe\,{\sc ii} and Ca\,{\sc ii}, when available.

\section{Comparison to BD+65\,1637}
\label{sec:bd65}

Comparing to the HBe star analyzed in Paper~I of this series, BD+65\,1637, also a B2e star, there are some similarities and differences among the keys findings. Similar to BD+65\,1637, it was possible to find in almost all cases synthetic line profiles that match the observed emission lines of the three HBe stars considered here. The exceptions are the He\,{\sc i} line of HD\,76534, for which we suggest a composite spectrum indicative of an unresolved binary, the Ca\,{\sc ii} line of HD\,216629, for which the observed line profile is wider than any of the model lines, and the He\,{\sc i} line of HD\,216629, which required a disk model to fit the observed line profile, as oppose a line of photospheric origin. In Paper~I, the individual line fit for the Ca\,{\sc ii} profile of BD+65\,1637 was much better, so the failure in the case of HD\,216629 does not seem to represent an inability of the present codes to model this line.

As with BD+65\,1637, it was not possible for any of the three HBe stars studied to find a single, global disk density model based on a single power-law that predicted emission line profiles in good accord with all observed lines. Therefore, as in Paper~I, it must be concluded that disk models with more complex density variations are required in order to be able to reproduce all the observed profiles with a unified model. 

It was also noted in Paper~I that BD+65\,1637 required higher density and lower power-law index models for the metal lines (Ca\,{\sc ii} \& Fe\,{\sc ii}) as compared to the Balmer lines. The trend is not so clear in the present work: HD\,76534 requires the Fe\,{\sc ii} lines to have higher $\rho_0$ and lower $n$ compared to the Balmer lines, whereas HD\,114981 requires the Fe\,{\sc ii} lines to have higher $\rho_0$ and higher $n$. The situation is even less clear for HD\,216629 where only H$\alpha$ disagrees with the location in the $(n,\log\rho_0$) plane selected by the metal lines, including Ca\,{\sc ii} in this case, and H$\beta$. 

Interestingly, the best global fit model for BD+65\,1637, excluding the Ca\,{\sc ii} lines, agrees with the single, global model found in the present work: $(n,\log\rho_0)=(3.0,-10)$. However, it must be restated that the global model line fits are, in general, much poorer than the individual line fits for all four stars. Finally, the disk size for BD+65\,1637 was found to be somewhat larger at $R_{\rm disk}=50\,R_*$ compared to the $25\,R_*$ required for the three stars of the present work.


\begin{figure*}
\centering
\includegraphics[width=0.99\textwidth]{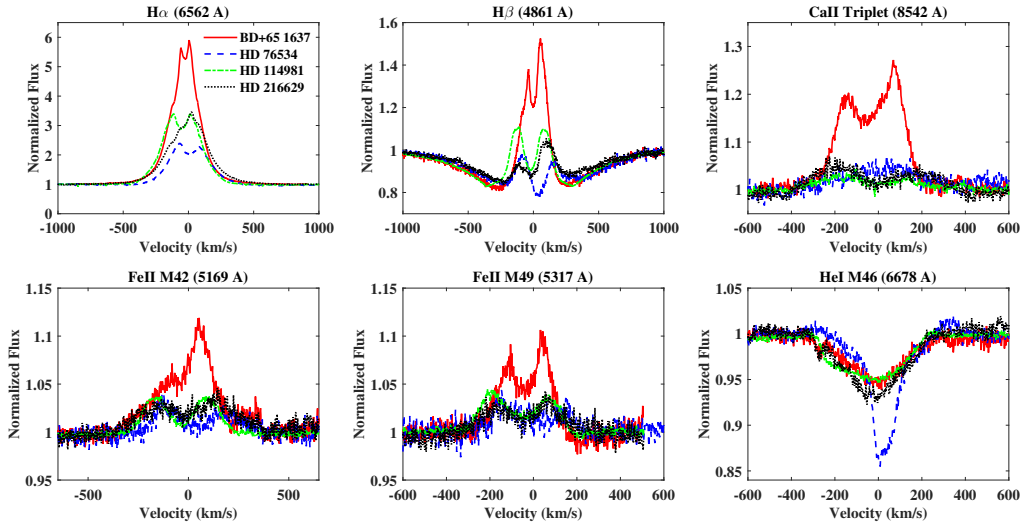}
\caption[An over-plot of line profiles for all the HB2e stars: BD+65\,1637, HD\,76534, HD\,114981 \& HD\,216629]{All the line profiles of all the stars considered in this work as well as in the previous work are overplotted. BD+65\,1637 studied in the previous work (Paper~I) is shown in red, while the three stars, HD\,76534, HD\,114981 and HD\,216629, studied here are shown in blue, green and black respectively.}
\label{fig:allstars_compared}
\end{figure*}

A plot of all the emission lines for all four stars is given in Figure~\ref{fig:allstars_compared} which shows how similar the emission line strengths and shapes are for the three HBe stars studied in the present work. BD+65\,1637 is seen to have significantly stronger emission lines, and this seems to be due to the larger size of its inner gaseous disk $(\sim 50\,R_*$) as compared to the three stars studies here ($\sim 25,R_*$). Among the He\,{\sc i} lines of the four stars, only the profile of HD\,76534 is clearly different, which has been very tentatively attributed to an unresolved contribution from a binary companion. 

As noted in Section~\ref{sec:intro}, the H$\alpha$ emission line of several HAeBe stars has been studied using spectro-interferometric observations taken with the VEGA instrument on the CHARA array. \cite{RP2010} studied the HAe star AB~Aur and modelled H$\alpha$ using a combination of magneto-centrifugal acceleration and a wind. Their model suggests that the H$\alpha$ emission arises from a disk of 2.3\,AU (and as close as 0.2\,AU from the star). Two other HAe stars, HD 179218 and HD 141569, were studied by~\cite{Mendigutia2016} and estimated to have H$\alpha$ emitting regions of size smaller than 0.21 and 0.12 AU respectively. Only one HBe star, MWC 361 (HD 200775), a known magnetic HBe star, has been studied using spectro-interferometric observations. \cite{Benisty2013} found that the H$\alpha$ line is emitted from disk beyond the regions generally inferred from the accretion process. They estimated the size of the H$\alpha$ emitting region to be 0.22\,AU. The values estimated for the H$\alpha$ emitting regions for the HAe stars may not be relevant to the three HBe stars considered in this study. However, the result of MWC~361 can be directly compared to that of the stars examind in this study. The individual fits to the H$\alpha$ line profile for these three stars range anywhere from 25\,R$_{*}$ (0.78 AU) to 100\,R$_{*}$ (3.1 AU). Given that observations for only one early-type Herbig Be star are available to compare to our results, it would be very valuable to have further interferometry of the early-type Herbig Be stars to extend the comparisons.

\section{Conclusions}
\label{sec:conclusions}

The key findings of this study of the inner gaseous disk of three HBe stars, HD\,76534, HD\,216629 and HD\,114981, are:

\begin{itemize}
    \item The emission line spectra of all three stars can be adequately reproduced, in terms of both line strength and shape, from a small ($\sim 25\,R_*$), geometrically thin, gaseous circumstellar disk heated solely by the available photoionizing radiation field from the photosphere of central B star.
    \item The equatorial density in the disks varies roughly as $10^{-10} \left(R_*/R\right)^3\,\rm g\,cm^{-3}$.
    \item The size and mass of the emitting disk are $R_{\rm disk}/R_*\approx\,25$ and $M_{\rm disk}/M_*\approx 10^{-9}$. 
    \item The models, however, are not successful in reproducing the line profiles of H\,{\sc i}, He\,{\sc i}, Ca\,{\sc ii} and Fe\,{\sc ii} simultaneously based on an equatorial disk density varying as a single power-law with radius. A possible resolution is to consider equatorial density distributions that vary in a more complex way. 
    \item The equatorial density distribution of the three stars is the same as found for the HBe B2 star BD+65\,1637 analyzed in Paper I, although their disks are approximately a factor of two smaller in outer radius.
\end{itemize}

We note that the current study focused exclusively on line emission from the inner gaseous disk; however, continuum emission is also expected, at least in principle, either in the near-IR from mostly hydrogen free-free emission or in the UV from hydrogen bound-free emission. It would be important to quantify the expected level of this emission from the disks above to see if this continuum emission can make a significant contribution to either (1) the near-IR excess of HBe stars, perhaps addressing why the HBe stars seem undersized compared to the predictions of the disk size -- luminosity correlation found by \citet{Mon02}, or (2) the UV excess seen in some HBe stars that does not seem to be able to be reproduced by magnetospheric accretion models \citep{Fairlamb2015}.

Finally, an important limitation of the current study is that the three HBe stars analyzed here plus BD+65\,1637 of Paper~I are all Group~III Herbig Ae/Be objects according to the \cite{Hillenbrand1992} classification, that is, objects with small infrared excesses resembling classical Be stars. We intend to extend our current analysis to Group~I and II objects that show unambiguous signs of dust in their infrared spectral energy distributions. We are also extending this analysis to both hotter and cooler HBe stars.

\vspace{1.5\baselineskip}

The authors would like to thank the anonymous referee for many helpful comments. This work is supported by the Canadian Natural Sciences and Engineering Research Council (NESRC) through Discovery Grants to T. A. A. Sigut and J. D. Landstreet.

\end{document}